\documentclass[prd,aps,tightenlines,nofootinbib,superscriptaddress]{revtex4}
\pdfoutput=1
\usepackage{epsfig}
\usepackage{amsmath}
\usepackage{graphicx}
 \usepackage{multirow}
\newcommand{\beq}{\begin{equation}}
\newcommand{\eeq}{\end{equation}}
\newcommand{\bqa}{\begin{eqnarray}}
\newcommand{\eqa}{\end{eqnarray}}

\begin{document}

\title{Model independent investigation of the $R_{J/\psi,\eta_c}$  and ratios of decay widths of semileptonic $B_c$ decays into a P-wave charmonium }

\author{Wei Wang$^2$,
Ruilin Zhu$^{1}$~\footnote{rlzhu@njnu.edu.cn} }
\affiliation{
$^1$ Department of Physics and Institute of Theoretical Physics,
Nanjing Normal University, Nanjing, Jiangsu 210023, China\\
$^2$SKLPPC, MOE Key Laboratory for Particle Physics, Astrophysics and Cosmology,  School of Physics and Astronomy, \\ Shanghai Jiao Tong University, Shanghai, 200240,   China }

\begin{abstract}
Experimental measurements of decay branching fractions of  semitaunic and semimuonic $B_c$   into $J/\psi$ has  challenged the lepton flavor  universality in   standard model with about  two standard deviations.  In this paper, we first investigate the unitary constraint on   form factors of   $B_c$ meson into $S$-wave and $P$-wave charmonium.  Such constraint  leads to  the exploration  of the $R_{J/\psi}$ and other ratios $R_{\eta_c}$, $R_{h_c}$, and $R_{\chi_{cJ}}$ in a model-independent way.  These results together with future experimental measurements    can   be  used  to explore the lepton flavor  universality in a more systematic way. In addition,  we point out that the helicity-dependent ratios $R^L_{J/\psi}$ and $R^\perp_{J/\psi}$ can  also provide complementary information.
\keywords{Semileptonic $B_c$ decays,
Lepton universality,
Unitary constraints.}

\end{abstract}

\maketitle

\section{Introduction}
Weak decays of heavy mesons play an important role in   testing  the standard model (SM) precisely and probing   new physics effects beyond SM.
In SM,  the  lepton interactions  are universal for all   three generations. Semileptonic decays of $B$ mesons induced by the $b\to c$   transition have revealed hints    lepton flavor non-university~\cite{Li:2018lxi}, which may indicate the presence of possible new physics (NP). To probe the NP effects in a more systematic way, more accurate and solid predictions of branching fractions   are highly  demanded.

Ratios of    semitaunic and semimuonic $\bar{B}$ decays into $D$ or $D^*$, i.e.  $R_D$ and $R_{D^*}$ have been measured by many experiments such as BABAR~\cite{Lees:2012xj}, Belle~\cite{Huschle:2015rga}, and LHCb~\cite{Aaij:2015yra}. These measurements
indicate the data of  $R_D$ and $R_{D^*}$  deviate from the SM predictions by about $3\sigma$.
Recently, the LHCb collaboration have reported the measurement of the ratio  of  the semitaunic and semimuonic $B^+_c$ decays into $J/\psi$,
i.e. $R_{J/\psi}$, using a sample of proton-proton collision data corresponding to $3.0fb^{-1}$ of  integrated luminosity~\cite{Aaij:2017tyk}. A deviation,  about $2\sigma$,  from SM is found in this measurement  by LHCb.

On the theoretical side, a few  schemes  are used  to study  the $B_c$ decays into a charmonium,  such as the perturbative QCD (PQCD) approach~\cite{Du:1988ws,Sun:2008ew,Wen-Fei:2013uea,Rui:2015iia,Rui:2014tpa,Liu:2018kuo,Xiao:2013lia,Rui:2018kqr,Rui:2017pre},  QCD sum rules (QCD SR)~\cite{Colangelo:1992cx,Kiselev:1993ea,Kiselev:1999sc,Azizi:2009ny}, light-cone sum rules (LCSR)~\cite{Huang:2007kb,Wang:2007fs},  quark model (QM)~\cite{Nobes:2000pm,Ebert:2003cn,Ivanov:2005fd,Ebert:2010zu,Hernandez:2006gt,Wang:2011jt,Tran:2018kuv},   light-front quark model (LFQM)~\cite{Wang:2008xt,Wang:2009mi,Ke:2013yka}, nonrelativistic QCD (NRQCD)~\cite{Chang:1992pt,Chang:2001vq,Chang:2001pm,Kiselev:2001zb,
 Bell:2005gw,Qiao:2011yz,Qiao:2012vt,Qiao:2012hp,Shen:2014msa,Zhu:2017lqu,Zhu:2017lwi} and others~\cite{Hsiao:2016pml}.
Uncertainties in  some of these schemes are hard to control especially  in the nonperturbative region and power corrections are expected important~\cite{Wang:2017jow}. At zero recoil, many of these theoretical investigations  become less solid and the extrapolations  must be used in order to access the full momentum dependence  of form factors.
On the lattice,  HPQCD collaboration have calculated  form factors for decays $B_c$ into the S-wave charmonium in the entire $q^2$ range, using a highly improved lattice quark action~\cite{Colquhoun:2016osw,Lytle:2016ixw}. These works make the analysis of semileptonic $B_c$ decays sitting on a firm ground.

In this paper, we will adopt a model-independent way to study the form factors of $B_c$ into the S-wave and P-wave charmonium in the entire phase space. Using the dispersion relation and crossing symmetry, we will give a new parametrization form expanded in a series of polynomials. The coefficients in the expansion can be obtained by fitting the data or the   Lattice simulations. At low recoil,  the heavy quark  limit reduces  form factors into a few Isgur-Wise functions, which can also be used to constrain the expansion. In the literature, the unitary constraint has been applied to study  $B$ meson decays into   light pseudoscalar and vector mesons, the charmed mesons~\cite{Singh:1977et,deRafael:1993ib,Boyd:1994tt,Neubert:1994vy,Boyd:1995cf,Boyd:1995sq,Caprini:1997mu,
Bourrely:2008za,
Caprini:2017ins,Caprini:2018jjm,Cohen:2018vhw,Huang:2018nnq}. Recently, this approach has been conducted  in $B_c$ into the S-wave charmonium  Refs.~\cite{zhu2017,Cohen:2018dgz,Murphy:2018sqg,Berns:2018vpl}. We extend this analysis to $B_c$ decays into all S-wave and P-wave charmonium in this work. Other studies on the $R_{J/\psi}$ can be found in Refs.~\cite{Dutta:2017xmj,Dutta:2017wpq,He:2017bft,Wei:2018vmk,Issadykov:2018myx,Davies:2018toj,Watanabe:2017mip}.

This paper is scheduled as follows. We give the form factors of $B_c$ decays into charmonia in Sec.~\ref{II}. These form factors will be reduced in heavy quark limit in Sec.~\ref{III}. We investigate the unitary constraints on form factors in Sec.~\ref{IV}. Semitaunic and semimuonic $B_c$ decays into other  $S$-wave and $P$-wave charmonia will also be studied and ratios for decay widths  will be presented.   In the last section, we give a summary of the paper.

\section{Form factors\label{II}}

Semileptonic decay amplitudes of   $B_c$ meson into charmonia   are characterized by  transition form factors, and  for the S-wave charmonia, they  are defined as
\begin{eqnarray}
\langle \eta_{c}(p)\vert J^\mu_V\vert B_{c}(P)\rangle
&=&f^{\eta_{c}}_{0}(q^{2})
\frac{m_{B_{c}}^{2}-m_{\eta_{c}}^{2}}{q^{2}}q^{\mu}+f^{\eta_{c}}_{+}(q^{2})(P^{\mu}+p^{\mu}-\frac{m_{B_{c}}^{2}-
m_{\eta_{c}}^{2}}{q^{2}}q^{\mu})\,,\label{ff1}
\\
\langle J/\psi(p,\varepsilon^{*})\vert J^\mu_V\vert
B_{c}(P)\rangle &=&-\frac{2
V^{ J/\psi}(q^{2})}{m_{B_{c}}+m_{J/\psi}}\epsilon^{\mu\nu\rho\sigma}
\varepsilon_{\nu}^{*}p_{\rho}P_{\sigma}\,,
\\
\langle J/\psi(p,\varepsilon^{*})\vert J^\mu_A\vert B_{c}(P)\rangle & =&-i[2
m_{J/\psi}A^{ J/\psi}_{0}(q^{2})\frac{\varepsilon^{*}\cdot q}{q^{2}}q^{\mu}+(m_{B_{c}}+m_{J/\psi})A^{ J/\psi}_{1}(q^{2})
(\varepsilon^{*\mu}-\frac{\varepsilon^{*}\cdot q}{q^{2}} q^{\mu})
\nonumber\\&&-A^{ J/\psi}_{2}(q^{2})\frac{\varepsilon^{*}\cdot
q}{m_{B_{c}}+m_{J/\psi}}(
P^{\mu}+p^{\mu}-\frac{m_{B_{c}}^{2}-m_{J/\psi}^{2}}{q^{2}}q^{\mu})]
\,,
\end{eqnarray}
and the form factors of the $B_c$ meson to P-wave charmonia are
\begin{eqnarray}
\langle h_{c}(p,\varepsilon^{*})\vert J^\mu_V\vert B_{c}(P)\rangle
&=&-i[2
m_{h_c}A^{h_c}_{0}(q^{2})\frac{\varepsilon^{*}\cdot q}{q^{2}}q^{\mu}
+(m_{B_{c}}+m_{h_c})A^{h_c}_{1}(q^{2})
(\varepsilon^{*\mu}-\frac{\varepsilon^{*}\cdot q}{q^{2}} q^{\mu})\nonumber\\&&~~~~~-A^{h_c}_{2}(q^{2})\frac{\varepsilon^{*}\cdot
q}{m_{B_{c}}+m_{h_c}}(
P^{\mu}+p^{\mu}-\frac{m_{B_{c}}^{2}-m_{h_c}^{2}}{q^{2}}q^{\mu})]\,,
\end{eqnarray}
\begin{eqnarray}
 && \langle h_c(p,\varepsilon^{*})\vert J^\mu_A \vert
B_{c}(P)\rangle =\frac{2
V^{h_c}(q^{2})}{m_{B_{c}}+m_{h_c}}\epsilon^{\mu\nu\rho\sigma}
\varepsilon_{\nu}^{*}p_{\rho}P_{\sigma}\,,
\end{eqnarray}
\begin{eqnarray}
&&\langle \chi_{c0}(p)\vert J^\mu_A \vert B_{c}(P)\rangle
=f^{\chi_{c0}}_{0}(q^{2})
\frac{m_{B_{c}}^{2}-m_{\chi_{c0}}^{2}}{q^{2}}q^{\mu}+f^{\chi_{c0}}_{+}(q^{2})(P^{\mu}+p^{\mu}-\frac{m_{B_{c}}^{2}-
m_{\chi_{c0}}^{2}}{q^{2}}q^{\mu})\,,\nonumber\\
\end{eqnarray}
\begin{eqnarray}
\langle \chi_{c1}(p,\varepsilon^{*})\vert J^\mu_V\vert B_{c}(P)\rangle
&=&-i[2
m_{\chi_{c1}}A^{\chi_{c1}}_{0}(q^{2})\frac{\varepsilon^{*}\cdot q}{q^{2}}q^{\mu}
+(m_{B_{c}}+m_{\chi_{c1}})A^{\chi_{c1}}_{1}(q^{2})
(\varepsilon^{*\mu}-\frac{\varepsilon^{*}\cdot q}{q^{2}} q^{\mu})\nonumber\\&&-A^{\chi_{c1}}_{2}(q^{2})\frac{\varepsilon^{*}\cdot
q}{m_{B_{c}}+m_{\chi_{c1}}}(
P^{\mu}+p^{\mu}-\frac{m_{B_{c}}^{2}-m_{\chi_{c1}}^{2}}{q^{2}}q^{\mu})]\,,
\end{eqnarray}
\begin{eqnarray}
 && \langle \chi_{c1}(p,\varepsilon^{*})\vert J^\mu_A\vert
B_{c}(P)\rangle =\frac{2
V^{\chi_{c1}}(q^{2})}{m_{B_{c}}+m_{\chi_{c1}}}\epsilon^{\mu\nu\rho\sigma}
\varepsilon_{\nu}^{*}p_{\rho}P_{\sigma},
\end{eqnarray}
\begin{eqnarray}
\langle \chi_{c2}(p,\varepsilon^{*})\vert J^\mu_A\vert B_{c}(P)\rangle
&=&[2
m_{\chi_{c2}}A^{\chi_{c2}}_{0}(q^{2})\frac{\varepsilon^{*\alpha\beta}q_\beta}{q^{2}}q^{\mu}
+(m_{B_{c}}+m_{\chi_{c2}})A^{\chi_{c2}}_{1}(q^{2})
(\varepsilon^{*\mu\alpha}-\frac{\varepsilon^{*\alpha\beta}q_\beta}{q^{2}} q^{\mu})\nonumber\\&&~-A^{\chi_{c2}}_{2}(q^{2})\frac{\varepsilon^{*\alpha\beta}
q_\beta}{m_{B_{c}}+m_{\chi_{c2}}}(
P^{\mu}+p^{\mu}-\frac{m_{B_{c}}^{2}-m_{\chi_{c2}}^{2}}{q^{2}}q^{\mu})
]\frac{-i P_{\alpha}}{m_{B_c}}\,,
\end{eqnarray}
\begin{eqnarray}
\langle \chi_{c2}(p,\varepsilon^{*})\vert J^\mu_V\vert
B_{c}(P)\rangle &=&\frac{2
V^{\chi_{c2}}(q^{2})}{m_{B_c}(m_{B_{c}}+m_{\chi_{c2}})}\epsilon^{\mu\nu\rho\sigma}
\varepsilon_{\nu\alpha}^{*}p_{\rho}P_{\sigma}P_{\alpha}\,.\label{ffchic2}
\end{eqnarray}
The vector and axial currents are defined as $J^\mu_V=\bar c \gamma^{\mu}b$ and $J^\mu_A=\bar c \gamma^{\mu}\gamma^{5}b$.
The momentum transfer is defined as $q=P-p$. If one defines $t_\pm\equiv m_{B_c}^2\pm m_{H}^2$ with the heavy quarkonium mass $m_H$,  the kinematics constraint is $0\leq q^2\leq t_-$.

\section{heavy quark effective theory\label{III}}

The transition  form factors at the large recoil   with $q^2\simeq 0$ might  be calculable  in  QCD factorization, however, when the $q^2$ approaches the $q^2_{\rm max}=t_-$, the perturbative expansion in $\alpha_s$ becomes less trustworthy. In this kinematics region,  the involved degrees of freedoms are either heavy quark or soft gluons, which can be handled in perturbative QCD  based on heavy quark effective theory (HQET).
In heavy quark limit, the heavy quark spin and flavor symmetries hold,  and thus decay constants and form factors are greatly simplified. In this section  we will analyze  the reduction of form factors in the heavy quark limit.

The four velocities of   $B_c$ and charmonia $H$ are denoted  as $v^\mu=P^\mu/m_{B_c}$ and ${v'}^\mu=p^\mu/m_{H}$, respectively.
The parameter $\omega\equiv v\cdot v'$ then becomes $(m_{B_c}^2+m_{H}^2-q^2)/(2m_{B_c}m_{H})$. The physic region for the parameter $
\omega$ is $1\leq\omega\leq (m_{B_c}^2+m_{H}^2)/(2m_{B_c}m_{H})$.

The outgoing  charm quark in the $b\to c$ transition is  at rest when $q^2=t_-$. The spectator charm quark is not involved in the hard scattering  in a short time at order of $1/m_W$.  It is convenient to investigate the symmetry for the $b\to c$ form factors in the  heavy quark limit.  The S-wave and P-wave charmonia can be classified into three multiplets. For S-wave charmonia, one has~\cite{Isgur:1990kf,Isgur:1990jg}
\begin{align}
H_{v'}  &=\frac{1+v'\!\!\!\!\slash}{2}[i{\psi}^\beta\gamma_\beta+{\eta_c}\gamma^5]\,.\label{eq:hqet1}
 \end{align}
 The P-wave charmonia can be merged into two multiplets by~\cite{Isgur:1990jf,Veseli:1995ep}
 \begin{align}
E_{v'}      &=\frac{1+v'\!\!\!\!\slash}{2}[\chi_{c0}+i{h_c}^\beta\gamma_\beta]\,,\\
F^\alpha_{v'}     &=i\frac{1+v'\!\!\!\!\slash}{2}\{\chi_{c2}^{\alpha\beta}\gamma_\beta-\sqrt{\frac{3}{2}}\chi_{c1}^\beta\gamma^5[g^\alpha_\beta
-\frac{1}{3}\gamma_\beta(\gamma^\alpha-{v'}^\alpha)]\} \,.\label{eq:hqet2}
 \end{align}
Form factors in Eqs.~(\ref{ff1}-\ref{ffchic2}) are parametrized by the QCD currents. In heavy quark limit, the $b\to c$ QCD transition operator can be matched onto the effective current $J^{'\mu}_V=\bar c_{v'} \gamma^{\mu}b_v$ and $J^{'\mu}_A=\bar c_{v'} \gamma^{\mu}\gamma^{5}b_v$, with the matching coefficient calculable in QCD perturbation theory.
Then  the form factors   are parametrized by
\begin{align}
\langle H^c(v')|\bar{c}_{v'}\Gamma^\mu b_v|B_c(v)\rangle  &={\rm Tr}[\xi\bar{H}^{c}_{v'}\Gamma^\mu \bar{H}^b_{v}] \,,\label{eq:trace1}
 \end{align}
where the parameter $\xi$ can be constructed using $v$ and $v'$, and $| H^c(v')\rangle$ can be one of state  in Eqs.~(\ref{eq:hqet1}-\ref{eq:hqet2}). The general form for $\xi$ can be written as
\begin{eqnarray}
\xi=\xi_0+\xi_1 v\!\!\!\slash+\xi_2 v'\!\!\!\!\slash+\xi_3 v\!\!\!\slash v'\!\!\!\!\slash
\end{eqnarray}
where the $\xi_i(i=0,1,2,3)$ are functions of $\omega$. The relations $v\!\!\!\slash H^b_v=H^b_v$ and $v\!\!\!\!\slash \bar{H}^c_{v'}=-\bar{H}^c_{v'}$
 indicate that all these terms are reduced into the first $\xi_i$. Thus one can write Eq.~(\ref{eq:trace1}) as
  \begin{align}
\langle H^c(v')|\bar{c}_{v'}\Gamma^\mu b_v|B_c(v)\rangle  &={\rm Tr}[\bar{H}^{c}_{v'}\Gamma^\mu \bar{H}^b_{v}]\xi_0(\omega) \,,\label{eq:trace2}
 \end{align}
where $\xi_0(\omega)=\xi_H(\omega)$ for $B_c$ to S-wave charmonium and  $\xi_0(\omega)=\xi_E(\omega),\xi_F(\omega)v_\alpha$ for $B_c$ to P-wave charmonium. At zero recoil   with $q^2=t_-$, we have the normalization  condition $\xi_{H}(\omega=1)=1$. Evaluating the above trace will lead to the form factors in heavy quark limit:
\begin{eqnarray}
\langle \eta_{c}(v')\vert \bar{c}_{v'}\gamma^\mu b_v \vert B_{c}(v)\rangle
&=&\xi_H(\omega)[v^\mu+{v'}^\mu]\,,\label{ff1-hqet}
\\
\langle J/\psi(v',\varepsilon^{*})\vert  \bar{c}_{v'}\gamma^\mu b_v \vert
B_{c}(v)\rangle &=&-\xi_H(\omega)\epsilon^{\mu\nu\rho\sigma}
\varepsilon_{\nu}^{*}{v'}_{\rho}v_{\sigma}\,,
\\
\langle J/\psi(v',\varepsilon^{*})\vert \bar{c}_{v'}\gamma^\mu\gamma^5 b_v \vert B_{c}(v)\rangle & =&-i\xi_H(\omega)[(1+\omega)
\varepsilon^{*\mu}-\varepsilon^{*}\cdot v {v'}^{\mu}]
\,,
\end{eqnarray}
\begin{eqnarray}
\langle h_{c}(v',\varepsilon^{*})\vert \bar{c}_{v'}\gamma^\mu b_v \vert B_{c}(v)\rangle
&=&i\xi_E(\omega)[(\omega-1)\varepsilon^{*\mu}-\varepsilon^{*}\cdot v {v'}^{\mu}]\,,\\
  \langle h_c(v',\varepsilon^{*})\vert  \bar{c}_{v'}\gamma^\mu\gamma^5 b_v \vert
B_{c}(v)\rangle & =&\xi_E(\omega)\epsilon^{\mu\nu\rho\sigma}
\varepsilon_{\nu}^{*}{v'}_{\rho}v_{\sigma}\,,\\
\langle \chi_{c0}(v')\vert  \bar{c}_{v'}\gamma^\mu \gamma^5 b_v \vert B_{c}(v)\rangle
&=&-\xi_E(\omega)[v^\mu-{v'}^\mu]\,,
\end{eqnarray}
\begin{eqnarray}
\langle \chi_{c1}(v',\varepsilon^{*})\vert  \bar{c}_{v'}\gamma^\mu b_v\vert B_{c}(v)\rangle
&=&\frac{i\xi_F(\omega)}{\sqrt{6}}[(\omega^2-1)\varepsilon^{*\mu}-\varepsilon^{*}\cdot v (3v^\mu-(\omega-2){v'}^{\mu})]\,,\\
 \langle \chi_{c1}(v',\varepsilon^{*}) \bar{c}_{v'}\gamma^\mu\gamma^5 b_v\vert
B_{c}(v)\rangle &=&\frac{(\omega+1)\xi_F(\omega)}{\sqrt{6}}\epsilon^{\mu\nu\rho\sigma}
\varepsilon_{\nu}^{*}{v'}_{\rho}v_{\sigma},\\
\langle \chi_{c2}(v',\varepsilon^{*})\vert  \bar{c}_{v'}\gamma^\mu\gamma^5 b_v \vert B_{c}(v)\rangle
&=&-i\xi_F(\omega)v_\alpha[(1+\omega)\varepsilon^{*\alpha\mu}-\varepsilon^{*\alpha\beta}v_\beta {v'}^{\mu}]\,,\\
\langle \chi_{c2}(v',\varepsilon^{*})\vert  \bar{c}_{v'}\gamma^\mu b_v\vert
B_{c}(v)\rangle &=&\xi_F(\omega)\epsilon^{\mu\nu\rho\sigma}
\varepsilon_{\alpha\nu}^{*}v^\alpha{v'}_{\rho}v_{\sigma}\,.\label{ffchic2-hqet}
\end{eqnarray}
Note that the hadronic states in QCD and the  effective theory differ by a factor $\sqrt{m_H}$ in leading order.
Thus the form factors defined in  Eqs.~(\ref{ff1}-\ref{ffchic2}) can be obtained in the heavy quark limit:
\begin{eqnarray}\label{HQET1}
f^{\eta_{c}}_{0}(\omega)
&=&\frac{(\omega +1) \xi _H(\omega) \sqrt{m_{B_c}} \sqrt{m_{\eta
   _c}}}{m_{B_c}+m_{\eta _c}}\,,\\
  f^{\eta_{c}}_{+}(\omega)
&=&\frac{\xi _H(\omega) \left(m_{B_c}+m_{\eta _c}\right)}{2 \sqrt{m_{B_c}} \sqrt{m_{\eta _c}}}\,,\\
V^{J/\psi}(\omega)
&=&\frac{\xi _H(\omega) \left(m_{B_c}+m_{J/\psi}\right)}{2 \sqrt{m_{J/\psi}} \sqrt{m_{B_c}}}\,,\\
 A^{J/\psi}_{0}(\omega)
&=&\frac{\xi _H(\omega) \left(m_{B_c}+m_{J/\psi}\right)}{2 \sqrt{m_{J/\psi}} \sqrt{m_{B_c}}}\,,\\
A^{J/\psi}_{1}(\omega)
&=&\frac{(\omega+1) \xi _H(\omega) \sqrt{m_{J/\psi}} \sqrt{m_{B_c}}}{m_{B_c}+m_{J/\psi}}\,,\\
A^{J/\psi}_{2}(\omega)
&=&\frac{\xi _H(\omega) \left(m_{B_c}+m_{J/\psi}\right)}{2 \sqrt{m_{J/\psi}} \sqrt{m_{B_c}}}\,.
\end{eqnarray}
\begin{eqnarray}
f^{\chi_{c0}}_{0}(\omega)
&=&\frac{(\omega -1) \xi _E(\omega) \sqrt{m_{B_c}} \sqrt{m_{\chi
   _{\text{c0}}}}}{m_{B_c}-m_{\chi _{\text{c0}}}}\,,\\
  f^{\chi_{c0}}_{+}(\omega)
&=&\frac{\xi _E(\omega) \left(m_{B_c}-m_{\chi _{\text{c0}}}\right)}{2 \sqrt{m_{B_c}} \sqrt{m_{\chi _{\text{c0}}}}}\,,\\
V^{h_c}(\omega)
&=&\frac{\xi _E(\omega)\left(m_{B_c}+m_{h_c}\right)}{2 \sqrt{m_{B_c}} \sqrt{m_{h_c}}}\,,\\
 A^{h_c}_{0}(\omega)
&=&\frac{\xi _E(\omega) \left(m_{B_c}-m_{h_c}\right)}{2 \sqrt{m_{B_c}}
   \sqrt{m_{h_c}}}\,,\\
A^{h_c}_{1}(\omega)
&=&-\frac{(\omega -1) \xi _E(\omega) \sqrt{m_{B_c}} \sqrt{m_{h_c}}}{m_{B_c}+m_{h_c}}\,,\\
A^{h_c}_{2}(\omega)
&=&-\frac{\xi _E(\omega) \left(m_{B_c}+m_{h_c}\right)}{2 \sqrt{m_{B_c}}
   \sqrt{m_{h_c}}}\,.
\end{eqnarray}
\begin{eqnarray}
V^{\chi_{c1}}(\omega)
&=&\frac{(\omega +1) \xi _F(\omega) \left(m_{B_c}+m_{\chi _{\text{c1}}}\right)}{2 \sqrt{6} \sqrt{m_{B_c}} \sqrt{m_{\chi _{\text{c1}}}}}\,,\\
 A^{\chi_{c1}}_{0}(\omega)
&=&-\frac{\xi _F(\omega)
   \left(m_{B_c}-m_{\chi _{\text{c1}}}\right) \left(3 m_{B_c}-(\omega -2) m_{\chi _{\text{c1}}}\right)}{2 \sqrt{6} \sqrt{m_{B_c}} m_{\chi
   _{\text{c1}}}^{3/2}}\,,\\
A^{\chi_{c1}}_{1}(\omega)
&=&-\frac{\left(\omega ^2-1\right) \xi _F(\omega) \sqrt{m_{B_c}} \sqrt{m_{\chi _{\text{c1}}}}}{\sqrt{6} \left(m_{B_c}+m_{\chi
   _{\text{c1}}}\right)}\,,\\
A^{\chi_{c1}}_{2}(\omega)
&=&-\frac{(\omega -5) \xi _F(\omega) \left(m_{B_c}+m_{\chi _{\text{c1}}}\right)}{2 \sqrt{6} \sqrt{m_{B_c}} \sqrt{m_{\chi _{\text{c1}}}}}\,,\\
   V^{\chi_{c2}}(\omega)
&=&\frac{\xi _F (\omega)\left(m_{B_c}+m_{\chi _{\text{c2}}}\right)}{2 \sqrt{m_{B_c}} \sqrt{m_{\chi _{\text{c2}}}}}\,,\\
 A^{\chi_{c2}}_{0}(\omega)
&=&\frac{\xi _F(\omega) \left(m_{B_c}+m_{\chi
   _{\text{c2}}}\right)}{2 \sqrt{m_{B_c}} \sqrt{m_{\chi _{\text{c2}}}}}\,,\\
A^{\chi_{c2}}_{1}(\omega)
&=&\frac{(\omega +1) \xi _F (\omega)\sqrt{m_{B_c}} \sqrt{m_{\chi _{\text{c2}}}}}{m_{B_c}+m_{\chi
   _{\text{c2}}}}\,,\\
A^{\chi_{c2}}_{2}(\omega)
&=&\frac{\xi _F(\omega) \left(m_{B_c}+m_{\chi _{\text{c2}}}\right)}{2 \sqrt{m_{B_c}} \sqrt{m_{\chi _{\text{c2}}}}}\,.\label{HQETn}
\end{eqnarray}
From the above formulae, one can see that these sixteen form factors for the $B_c$ into S-wave and P-wave charmonia can be obtained by three
universal Isgur-Wise functions~\cite{Isgur:1990kf}, i.e. $\xi _H(\omega)$, $\xi _E(\omega)$ and $\xi _F(\omega)$.

\section{Unitatity constraints\label{IV}}

\subsection{Dispersion relation}
In the following, we will consider the constraints on form factors from unitarity, completeness and causality. The unitarity bound
of form factors via dispersion
relation and quark-hadron duality will be obtained.
Let us first consider the  relevant flavour-changing vector  and axial-vector currents
\begin{align}
 j^{\mu}_V      &= \bar c\gamma ^\mu  b  \,,
&
j^{\mu}_A &= \bar c\gamma ^\mu   \gamma^5  b \,.\label{eq:current}
 \end{align}
In QCD the two-point correlation function of two currents is defined as
\begin{eqnarray}
\Pi^{\mu \nu}(q^2) & = & i \, \int d^4 x \, e^{i \, q \cdot x}
 \left\langle 0 \right|{\rm T} \, j^{\mu}(x) \, j^{\dag\,\nu}(0)\left| 0 \right\rangle
 \nonumber\\
 &=&(\frac{q^\mu  q^\nu}{q^2}-g^{\mu \nu })\Pi_{T}(q^2)+\frac{q^\mu  q^\nu}{q^2}\Pi_{L}(q^2)  \,,
\label{eq:correlator}
\end{eqnarray}
which  can be evaluated by the operator product expansion (OPE) approach.
The corresponding longitudinal Lorenz scalar $\Pi_{L}(q^2)$ and transverse Lorenz scalar $\Pi_{T}(q^2)$ can be obtained by
 \begin{equation}
\Pi_{I}(q^2) = P_{\mu \nu ,I}(q^2) \, \Pi^{\mu \nu }(q^2)\,, \qquad
 \mbox{\small ($I=L,T$),}
\end{equation}
with the longitudinal and transverse helicity projectors,
 \begin{equation}
 P _{\mu \nu,L }(q^2)  = \frac{{q{}_\mu q_\nu  }}{{q^2}}
\,, \qquad
 P _{\mu \nu ,T}(q^2)  = \frac{1}{(D-1)}\left( \frac{q_\mu  q_\nu}{q^2 }-g_{\mu \nu } \right)
\,.
 \end{equation}
 $\Pi _{I}(q^2)$ is an analytic function
and it satisfies the dispersion relation
\begin{equation} \label{eq:chidef}
\Pi_{I}(q^2)= \frac{1}{\pi } \, \int\limits_0^\infty
dt \, \frac{\mathrm{Im} \, \Pi_I (t)}{ {t - q^2 }  }\,.
\end{equation}
The n-th subtracted dispersion relation is
\begin{equation} \label{eq:chindef}
\chi_I(n,Q_0^2) =\frac{1}{n!}\left. {\frac{d^n\Pi_{I}(q^2)}{{{dq^2}^n}}} \right|_{q^2  = -Q_0^2}= \frac{1}{\pi } \, \int\limits_0^\infty
dt \, \frac{\mathrm{Im} \, \Pi_I (t)}{\left( {t +Q_0^2 } \right)^{n+1} }\,.
\end{equation}

\subsection{Resonance contributions}

Using the dispersion relation in Eqs.~(\ref{eq:chidef}-\ref{eq:chindef}), we  need to calculate the imaginary part of the correlators.
The  imaginary part  $\mathrm{Im}\, \Pi_{I}(q^2)$ can be  obtained through inserting complete basis with all hadronic states
with allowed quantum numbers. The complete set of hadronic states $X$ have the identical quantum numbers as the vector or axial-vector current $j_\mu$.

For a particular choice of intermediate state $X= B V$, where $BV$ may be either $BD^*$ or $B_c J/\psi$,  we define
 \begin{equation}
\mathrm{Im} \, \Pi _{I}^{BV}(q^2)
 =\frac{1}{2}\, \int {\frac{d^3 p_B}{(2\pi)^32E_B} \, \frac{d^3 p_V}{(2\pi)^32E_V} \, (2\pi)^4\delta^4(q - p_B  - p_V)} \, P_{\mu \nu,I } \,
\left\langle 0 \right|j^\mu  \left| B V \right\rangle
\left\langle  B V \right|j^{\nu\dag}  \left|0 \right\rangle\,,
\end{equation}
where $q^2\geq m_B^2-m_V^2$.
Because the inserted hadronic states are not complete, it results in the inequality
\begin{equation}\label{eq:unit}
{\mathop{\rm Im}\nolimits} \Pi^{B V}_{I} (t) \leq
{\mathop{\rm Im}\nolimits} \Pi_{I} (t) \,.
\end{equation}
In principle,  we need to consider enough hadronic states to close to the bound. However, there is no conflict to write the expression as in Eq.~(\ref{eq:unit}).

One can use the crossing symmetry to relate the matrix elements $\left\langle 0  \right|j^\mu  \left| BV \right\rangle$ to $\left\langle B \right|j^\mu  \left| V \right\rangle$,
 replacing  $p_V$ in $\left\langle B \right|j^\mu   \left| V \right\rangle$ to $-p_V$. Then we obtain the following identity
\begin{eqnarray}
\nonumber P _{\mu \nu,T } \,
 \langle 0|   j^\mu_V  | BV\rangle
\langle B V|j_V ^{\nu\dag}|0\rangle
&=&\frac{\lambda}{3 q^2}\, \left|\mathcal{B}_{V}\right|^2
\,, \\
\nonumber P _{\mu \nu, T } \,
 \langle 0|   j^\mu_A  | BV\rangle
\langle B V|j_A ^{\nu \dag}|0\rangle
&=&\frac{\lambda}{3q^2 }\,\sum_{i=1}^2 \left|\mathcal{B}_{Ai}\right|^2 \,,
 \\
 P _{\mu \nu, L} \,
 \langle 0|   j^\mu_A  | BV\rangle
\langle B V|j_A ^{\nu\dag}|0\rangle
&=&\frac{\lambda}{3 q^2}\, \left|\mathcal{B}_{A0}\right|^2
\,,
\label{BFF2}
\end{eqnarray}
where
\begin{align}
 \mathcal{B}_{V}(q^2)&= \frac{\sqrt{2q^2}  }{m_B+m_V}V(q^2) \,,
\cr
\mathcal{B}_{A0}(q^2) & =\sqrt{3} A_0(q^2) \,,
\cr
\mathcal{B}_{A1}(q^2)&= \frac{(m_B+m_V)^2 \, (m_B^2-m_V^2-q^2) \, A_1(q^2)
 -\lambda \, A_2(q^2)}{2 m_V \sqrt{\lambda} \, (m_B+m_V)} \,,
\cr
\mathcal{B}_{A2}(q^2)&
=\frac{\sqrt{2 \, q^2 } \, (m_B+m_V)}{\sqrt{\lambda }} \, A_1(q^2) \,,\label{BFF3}
\end{align}
and
\begin{align}
\lambda &= \left( (m_B-m_V)^2-q^2\right) \left( (m_B+m_V)^2-q^2\right),
\label{eq:lambdadef}
\end{align}
which satisfies the identity $\lambda\equiv (t_- - q^2)(t_+ - q^2)$ in the case of $B_c$ and $J/\psi$.

One  can now express $\mathrm{Im} \, \Pi^{B V} _{I,i} $ in a compact form,
\begin{eqnarray}
\mathrm{Im} \, \Pi^{B V} _{I,i}&=&
\frac{1}{2}\, \int {\frac{d^3 p_B}{(2\pi)^32E_B} \, \frac{d^3 p_V}{(2\pi)^32E_V} \, (2\pi)^4\delta^4(q - p_B  - p_V)} \, \frac{\lambda}{3q^2} \left|A_{I,i}^{V}\right|^2\nonumber\\
&=&\frac{1}{48\pi} \,
\frac{\lambda^{3/2}}{q^{4}}
\left|A_{I,i}^{V}\right|^2 \,,\label{eq:egBM}
\end{eqnarray}
where the $\left|A_{I,i}^V\right|^2$ can be extracted from Eq.~(\ref{BFF2}),
\begin{align}
 \left|A_{T,V}^{V}\right|^2 &=\left|\mathcal{B}_{V}\right|^2 \,,
 \qquad
 \left|A_{T,A}^{V}\right|^2 =
  \sum_{i=1}^2 \left|\mathcal{B}_{Ai}\right|^2 \,,
\qquad
 \left|A_{L,A}^{V}\right|^2 =   \left|\mathcal{B}_{A0}\right|^2 \,.
\label{eq:Afunc1}
\end{align}

For a particular choice of intermediate state $X= B P$ where the $BP$ may be either $BD$ or $B_c \eta_c$,  we have
\begin{eqnarray}
\nonumber P _{\mu \nu,T } \,
 \langle 0|   j^\mu_V  | BP\rangle
\langle B P|j_V ^{\nu\dag}|0\rangle
&=&\frac{\lambda}{3 q^2}\, \left|\mathcal{B}_{f_+}\right|^2
\,, \\
\nonumber P _{\mu \nu,L  } \,
 \langle 0|   j^\mu_V  | BP\rangle
\langle B P|j_V ^{\nu \dag}|0\rangle
&=&\frac{\lambda}{3q^2 }\, \left|\mathcal{B}_{f_0}\right|^2 \,,
\label{BPF2}
\end{eqnarray}
where
\begin{align}
 \mathcal{B}_{f_+}(q^2)&=f_+(q^2) \,,
\cr
\mathcal{B}_{f_0}(q^2) & =\frac{\sqrt{3}(m_B^2-m_P^2 )}{\sqrt{\lambda}}f_0(q^2) \,.
\label{BPF3}
\end{align}
One  obtains  $\mathrm{Im} \, \Pi^{B P} _{I,V} $ in a compact form
\begin{eqnarray}
\mathrm{Im} \, \Pi^{B P} _{I,V}&=&
\frac{1}{2}\, \int {\frac{d^3 p_B}{(2\pi)^32E_B} \, \frac{d^3 p_P}{(2\pi)^32E_P} \, (2\pi)^4\delta^4(q - p_B  - p_P)} \, \frac{\lambda}{3q^2} \left|A_{I,V}^{P}\right|^2\nonumber\\
&=&\frac{1}{48\pi} \,
\frac{\lambda^{3/2}}{q^{4}}
\left|A_{I,V}^{P}\right|^2 \,,\label{eq:egBP}
\end{eqnarray}
where the $\left|A_{I,V}^P\right|^2$ can be extracted from Eq.~(\ref{BPF2}),
\begin{align}
 \left|A_{T,V}^{P}\right|^2 &=\left|\mathcal{B}_{f_+}\right|^2 \,,
\qquad
 \left|A_{L,V}^{P}\right|^2 =   \left|\mathcal{B}_{f_0}\right|^2 \,.
\label{eq:Afunc2}
\end{align}

Similarly, for the  axial vector and scalar  states, we have
\begin{align}
 \left|A_{T,A}^{A}\right|^2 &=\left|\mathcal{B}_{V0}\right|^2 \,,
 \qquad
 \left|A_{T,V}^{A}\right|^2 =
  \sum_{i=1}^2 \left|\mathcal{B}_{Ai}\right|^2 \,,
\qquad
 \left|A_{L,V}^{A}\right|^2 =   \left|\mathcal{B}_{A0}\right|^2 \,.
\label{eq:Afunc3}
\end{align}
and
\begin{align}
 \left|A_{T,A}^{S}\right|^2 &=\left|\mathcal{B}_{f_+}\right|^2 \,,
\qquad
 \left|A_{L,A}^{S}\right|^2 =   \left|\mathcal{B}_{f_0}\right|^2 \,.
\label{eq:Afunc4}
\end{align}

For the tensor state, one has
\begin{eqnarray}
\nonumber P _{\mu \nu,T } \,
 \langle 0|   j^\mu_V  | BT\rangle
\langle B T|j_V ^{\nu\dag}|0\rangle
&=&\frac{\lambda}{3 q^2}\, \left|\mathcal{B}^T_{V}\right|^2
\,, \\
\nonumber P _{\mu \nu, T } \,
 \langle 0|   j^\mu_A  | BT\rangle
\langle B T|j_A ^{\nu \dag}|0\rangle
&=&\frac{\lambda}{3q^2 }\,\sum_{i=1}^2 \left|\mathcal{B}^T_{Ai}\right|^2 \,,
 \\
 P _{\mu \nu, L} \,
 \langle 0|   j^\mu_A  | BT\rangle
\langle B T|j_A ^{\nu\dag}|0\rangle
&=&\frac{\lambda}{3 q^2}\, \left|\mathcal{B}^T_{A0}\right|^2
\,,
\label{BTF2}
\end{eqnarray}
where
\begin{align}
 \mathcal{B}^T_{V}(q^2)&= \frac{\sqrt{\lambda q^2}  }{2m_Bm_T(m_B+m_T)}V(q^2) \,,
\cr
\mathcal{B}^T_{A0}(q^2) & =\frac{\sqrt{\lambda }}{ \sqrt{2}m_B m_T } A_0(q^2) \,,
\cr
\mathcal{B}^T_{A1}(q^2)&= \frac{(m_B+m_T)^2 \, (m_B^2-m_T^2-q^2) \, A_1(q^2)
 -\lambda \, A_2(q^2)}{2  \sqrt{6 } \,m_B m_T^2 (m_B+m_T)} \,,
\cr
\mathcal{B}^T_{A2}(q^2)&
=\frac{ (m_B+m_T)\sqrt{q^2}}{2m_B m_T} \, A_1(q^2) \,. \label{BTF3}
\end{align}

The $\mathrm{Im} \, \Pi^{B T} _{I,i} $ becomes
\begin{eqnarray}
\mathrm{Im} \, \Pi^{B T} _{I,i}&=&
\frac{1}{2}\, \int {\frac{d^3 p_B}{(2\pi)^32E_B} \, \frac{d^3 p_T}{(2\pi)^32E_T} \,
(2\pi)^4\delta^4(q - p_B  - p_T)} \, \frac{\lambda}{3q^2} \left|A_{I,i}^{T}\right|^2\nonumber\\
&=&\frac{1}{48\pi} \,
\frac{\lambda^{3/2}}{q^{4}}
\left|A_{I,i}^{T}\right|^2 \,,\label{eq:egBT}
\end{eqnarray}
where the $\left|A_{I,i}^T\right|^2$ can be extracted from Eq.~(\ref{BTF2}),
\begin{align}
 \left|A_{T,V}^{T}\right|^2 &=\left|\mathcal{B}^T_{V}\right|^2 \,,
 \qquad
 \left|A_{T,A}^{T}\right|^2 =
  \sum_{i=1}^2 \left|\mathcal{B}^T_{Ai}\right|^2 \,,
\qquad
 \left|A_{L,A}^{T}\right|^2 =   \left|\mathcal{B}^T_{A0}\right|^2 \,.
\label{eq:Afunc5}
\end{align}

\subsection{OPE for the two-point correlation function}

The two-point correlation function defined in Eq.~(\ref{eq:correlator}) can be calculated using  OPE. The expression has  been given as~\cite{Shifman:1978bx,Shifman:1978by,Novikov:1980uj}
 \begin{eqnarray}
i \int dx \, e^{i \, q \cdot x} \,
\left\langle 0 \right|{\rm T} \, j^{\mu}(x) \, j^{\dag\,\nu}(0)\left| 0 \right\rangle
&=&(\frac{q^\mu  q^\nu}{q^2}-g^{\mu \nu })\sum\limits_{n = 1}^\infty\, C_{T,n}(q) \, \left\langle 0 \right|:{\cal O}_n(0):\left| 0 \right\rangle\,\nonumber\\&&+\frac{q^\mu  q^\nu}{q^2}\sum\limits_{n = 1}^\infty\, C_{L,n}(q) \, \left\langle 0 \right|:{\cal O}_n(0):\left| 0 \right\rangle \,,
\end{eqnarray}
where the local operators ${\cal O}_n(0)$ are constructed by quark and gluon fields. As for the flavor changing double-heavy currents defined in Eq.~(\ref{eq:current}), the first two dominant contributions come from the unit operator and the gluon vacuum  condensate $\displaystyle\langle (\alpha_s/\pi) \, G^2\rangle$. Here $C_{I,n}(q)$ are the related short-distance perturbative coefficients~\cite{Jamin:1992se}. Results for n-th subtracted scalars $\chi_I(n,Q_0^2=0)$ can be found in Eqs. (4.2-4.3) and (4.8-4.9) in Ref.~\cite{Boyd:1997kz}. We have
 \begin{eqnarray}
\chi^V_L(n=1)\bigg|_{r=0.286}
&=& 4.48\times 10^{-3} \left(1+1.34 \alpha_s-6.50\times10^{-4}\left(\frac{4.9\mathrm{GeV}}{m_b}\right)^4\frac{\displaystyle\langle \alpha_s G^2/\pi\rangle}{0.02\mathrm{GeV}^4}\right),\nonumber\\
\chi^A_L(n=1)\bigg|_{r=0.286}
&=& 2.09\times 10^{-2} \left(1+0.62 \alpha_s+2.87\times10^{-4}\left(\frac{4.9\mathrm{GeV}}{m_b}\right)^4\frac{\displaystyle\langle \alpha_s G^2/\pi\rangle}{0.02\mathrm{GeV}^4}\right),~~~
\end{eqnarray}
and
 \begin{eqnarray}
\chi^V_T(n=2)\bigg|_{r=0.286}
&=& \frac{9.94\times 10^{-3} }{m_b^2}\left(1+1.38 \alpha_s-8.69\times10^{-6}\left(\frac{4.9\mathrm{GeV}}{m_b}\right)^4\frac{\displaystyle\langle \alpha_s G^2/\pi\rangle}{0.02\mathrm{GeV}^4}\right),\nonumber\\
\chi^A_T(n=2)\bigg|_{r=0.286}
&=& \frac{6.10\times 10^{-3} }{m_b^2}\left(1+1.32 \alpha_s-8.40\times10^{-4}\left(\frac{4.9\mathrm{GeV}}{m_b}\right)^4\frac{\displaystyle\langle \alpha_s G^2/\pi\rangle}{0.02\mathrm{GeV}^4}\right),~~~~\end{eqnarray}
where $r=m_c/m_b$. In the above expression, we have seen that the gluon condensate contributions  are trivial in the correlator  of double heavy vector
and axial-vector currents.

\subsection{Bounds on expansion coefficients}

Using Eq.~(\ref{eq:unit}) and quark-hadron duality, one derives
 \begin{equation}
 \frac{1}{\pi } \,
 \int\limits_0^\infty dt \,
 \frac{\mathrm{Im} \, \Pi^{BH}_{I,X} (t)}{\left(t - q^2 \right)^{n+1} }\bigg|_{q^2=-Q_0^2}
=
 \frac{1}{\pi } \,
 \int\limits_{t_+}^\infty dt \,
 \frac{\lambda^{3/2}(t)}{48\pi \, t^{2}(t+Q_0^2)^{n+1}} \left| A_{I,X}(t) \right|^2 \leq
\chi^{X}_{I}(n,Q_0^2)\,,
\label{eq:unit2}
  \end{equation}
where $\chi^{X}_{X} \equiv \chi^{X}_{I,\rm OPE}$.
One can  choose
$Q_0^2=0\ll t_+$.

In Eq.~(\ref{eq:unit2}), we have $t\geq t_+$. In order to map the $t$-plane to the unit disk, one can define a parameter $z(t)$
\begin{equation}
\label{eq:zdef}
z(t)\equiv z(t,t_0)= \frac{ \sqrt{t_+-t}- \sqrt{t_+-t_0}}{\sqrt{t_+-t}+ \sqrt{t_+-t_0}} \,,
\end{equation}
where the free parameter $t_0$ satisfies  $0 \leq t_0<t_-$,
which can be optimised to reduce the maximum value of $|z(t)|$ in
the physical form factor range~\cite{Bharucha:2010im}
\begin{equation}
 t_0 \big|_{\rm opt.} =  t_+  - \sqrt{t_+(t_+-t_-)}
\,.
\label{eq:t0opt}
\end{equation}
Then the region  $0 \leq t<t_+$ can be mapped onto the unit disk $|z(t)|<1$, while the physical pair-production  region $t\geq t_+$ onto the unit circle $|z(t)|=1$.
Expressed by $z$, the inequality could be written in the form
\begin{align}
\label{eq:phidet}
\frac{1}{2\pi i } \,
\oint &\frac{dz}{z} \, |\phi_I^X (z)\, A _{I,X}(z)|^2 \leq 1
\quad \Leftrightarrow \quad
\frac{1}{\pi} \,
\int_{t_+}^\infty\frac{dt}{t-t_0} \, \sqrt{\frac{t_+-t_0}{t-t_+}}
\, |\phi_I^X(t) \, A_{I,X}(t)|^2 \leq 1
\,,
\end{align}
here the auxiliary function $\phi_I^X(t)$ can be obtained by comparing
(\ref{eq:phidet}) and (\ref{eq:unit2}),
\begin{equation}\label{eq:bound2a}
 |\phi_{I}^X(t)|^2= \frac{1}{48 \pi \, \chi^{X}_I(n)}
\, \frac{(t-t_+)^2}{(t_+ -t_0)^{1/2}} \, \frac{(t-t_-)^{3/2}}{t^{n+2}}
 \, \frac{t-t_0}{t} \,.
\end{equation}
The form factors $A_I^X(t)$ then can be generally expanded as~\cite{Bharucha:2010im}
\begin{equation}
A_{I,X}(t) = \frac{(\sqrt{-z(t,0)})^m
(\sqrt{z(t,t_-)})^l }{B(t) \, \phi_I^X(t)} \, \sum_{n=0}^\infty \alpha_n \, z^n\,,
\label{eq:ssepar}
\end{equation}
where the factors ${(\sqrt{-z(t,0)})^m}$ and
${(\sqrt{z(t,t_-)})^l}$ have been included  to take into account the poles at $t=0$ and $t=t_-$~\cite{Hill:2006ub}, and  the module of which
equals to 1 in the pair-production region.
The function  $B(t)$ is a Blaschke factor with ${B(t) = \prod_i z(t,m_{R_i}^2)}$,
representing resonances poles with masses $m^2_{R_i}\leq t_+$, and
satisfying $|B(t)| = 1$ in the pair-production region.
Following the same procedure to consider the poles at $t=0$ and $t=t_-$, we get
\begin{equation}
\phi_I^X(t)=\sqrt{\frac{1}{48\pi\chi^{X}_{I}(n)}}
\, \frac{(t-t_+)}{(t_+-t_0)^{1/4}}
\left(\frac{z(t,0)}{-t}\right)^{(n+3)/2}
\left(\frac{z(t,t_0)}{t_0-t}\right)^{-1/2}
\left(\frac{z(t,t_-)}{t_--t}\right)^{-3/4}\,.
\end{equation}
Using the expressions in Eqs. (\ref{eq:ssepar}) into (\ref{eq:phidet}),
and the inequality $|z(t,t_0)|\leq1$, $|z(t,m_R^2)|\leq1$ and $|z(t,0)|\leq1$, we obtain
the bound on the coefficients $\alpha_n$:
\begin{equation}
\sum_{n=0}^\infty \alpha_n^2<1 \,.
\end{equation}

\subsection{Parametrization  }

Using the series in Eq.~\eqref{eq:ssepar}, one can establish the parametrization of  the form factors $f_0$, $f_+$, $V$ and $A_i$, and the generic formula is given as
 \begin{equation}
\label{z-Se}
  F_i(t) = \frac{1}{B(t) \, \phi_i(t)} \, \sum_{k} \alpha^i_k \, z^k(t) \,.
 \end{equation}
For the $B_c$ meson decays to $J/\psi$, we denote $F_1(t)=V(t)$, $F_2(t)=A_0(t)$, $F_3(t)=A_1(t)$, and we also define two new form factors
 \begin{eqnarray}
  F_4(t) = \frac{\left(t_+-t\right)\left(\frac{z(t,t_-)}{t_--t}\right)^{-1} \, A_2(t)-(m_B+m_V)^2 \, (m_B^2-m_V^2-t) \, A_1(t)
 }{2 m_V (m_B+m_V)^3} \,,\nonumber\\
  F^\prime_4(t) = \frac{\lambda \, A_2(t)-(m_B+m_V)^2 \, (m_B^2-m_V^2-t) \, A_1(t)
 }{2 m_V (m_B+m_V)^3} \,.
 \end{eqnarray}
 The corresponding functions $\phi_i(t)$ are
 \begin{eqnarray}
\phi_1(t)&=&\frac{m_B+m_V}{\sqrt{2\chi^{V}_{T}(n)}}
\, \left(\frac{z(t,0)}{-t}\right)^{1/2}\phi_0(t)\,,\label{FV1}\\
\phi_2(t)&=&\frac{1}{\sqrt{3\chi^{A}_{L}(n)}}
\, \phi_0(t)\,,\\
\phi_3(t)&=&\frac{\left(t_+-t\right)^{1/2}}{\sqrt{2\chi^{A}_{T}(n)}(m_B+m_V)}
\, \left(\frac{z(t,0)}{-t}\right)^{1/2}\left(\frac{z(t,t_-)}{t_--t}\right)^{-1/2}\phi_0(t)\,,\\
\phi^{(\prime)}_4(t)&=&\frac{\left(t_+-t\right)^{1/2}}{\sqrt{\chi^{A}_{T}(n)}(m_B+m_V)^2}
\, \left(\frac{z(t,t_-)}{t_--t}\right)^{-1/2}\phi_0(t)\,,\label{FV4}
\end{eqnarray}
with
 \begin{equation}
\phi_0(t)=\sqrt{\frac{1}{48\pi}}
\, \frac{(t-t_+)}{(t_+-t_0)^{1/4}}
\left(\frac{z(t,0)}{-t}\right)^{(n+3)/2}
\left(\frac{z(t,t_0)}{t_0-t}\right)^{-1/2}
\left(\frac{z(t,t_-)}{t_--t}\right)^{-3/4}\,,
\end{equation}
where the relation is hold by $\phi_0(t)=\phi_I^X(t)\sqrt{\chi^{X}_{I}(n)}$.
The Blaschke factor ${B(t) = \prod_i z(t,m_{R_i}^2)}$,
representing poles due to sub-threshold resonances of masses $m_{R_i}$. There are not enough data
to determine the structure of $B(t)$. Fortunately, for $b\to c$ transition, the masses of the relevant $B_c$-type resonances can be estimated from potential model or Lattice QCD. Using the potential model expectation~\cite{Ebert:2011jc}, we get
\begin{eqnarray}
B^V(t)&=&z(t,6.333^2)z(t,6.882^2)z(t,7.021^2)z(t,7.258^2)z(t,7.392^2)\nonumber\\&&z(t,7.609^2)z(t,7.733^2)z(t,7.947^2)\,,\nonumber\\
B^A(t)&=&z(t,6.743^2)z(t,6.750^2)z(t,7.134^2)z(t,7.147^2)z(t,7.500^2)\nonumber\\&&z(t,7.510^2)z(t,7.844^2)z(t,7.853^2)\,.
\end{eqnarray}

For the $B_c$ meson decays to $\eta_c$, we denote $F^P_1(t)=f_+(t)$  and  $F^P_2(t)=f_0(t)$.
 The corresponding functions $\phi^P_i(t)$ are
 \begin{eqnarray}
\phi^P_1(t)&=&\frac{1}{\sqrt{\chi^{V}_{T}(n)}}
\, \phi_0(t)\,,\label{FP1}\\
\phi^P_2(t)&=&\frac{\left(t_+-t\right)^{1/2}}{\sqrt{3\chi^{V}_{L}(n)}(m^2_B-m_P^2)}\left(\frac{z(t,t_-)}{t_--t}\right)^{-1/2}
\, \phi_0(t)\,,\label{FP2}
\end{eqnarray}

Similarly, one can easily get the formulae when the final charmonium is $h_c$, $\chi_{c1}$ or $\chi_{c0}$. For $B_c$ decays to $\chi_{c2}$,
we denote $F^T_1(t)=V(t)$, $F^T_2(t)=A_0(t)$, $F^T_3(t)=A_1(t)$, and we also define two new form factors
 \begin{eqnarray}
  F^T_4(t) = \frac{\left(t_+-t\right)\left(\frac{z(t,t_-)}{t_--t}\right) \, A_2(t)-(m_B+m_T)^2 \, (m_B^2-m_T^2-t) \, A_1(t)
 }{2\sqrt{6} m_B m_T^2 (m_B+m_T)} \,,\nonumber\\
  F^{T\prime}_4(t) =  \frac{\lambda\, A_2(t)-(m_B+m_T)^2 \, (m_B^2-m_T^2-t) \, A_1(t)
 }{2\sqrt{6} m_B m_T^2 (m_B+m_T)}  \,.
 \end{eqnarray}
 The corresponding functions $\phi^T_i(t)$ are
 \begin{eqnarray}
\phi^T_1(t)&=&\frac{2m_B m_T(m_B+m_T)}{\sqrt{\chi^{V}_{T}(n)}}
\,\left(\frac{z(t,0)}{-t}\right)^{1/2}\left(t_+-t\right)^{-1/2} \left(\frac{z(t,t_-)}{t_--t}\right)^{1/2}\phi_0(t)\,,\label{FT1}\\
\phi^T_2(t)&=&\frac{\sqrt{2}m_B m_T}{\sqrt{\chi^{A}_{L}(n)}}
\,  \left(\frac{z(t,t_-)}{t_--t}\right)^{-1/2}\phi_0(t)\,,\\
\phi^T_3(t)&=&\frac{1}{\sqrt{\chi^{A}_{T}(n)}}
\, \phi_0(t)\,,\\
\phi^{T(\prime)}_4(t)&=&\frac{\left(t_+-t\right)^{1/2}}{\sqrt{\chi^{A}_{T}(n)}}
\, \left(\frac{z(t,t_-)}{t_--t}\right)^{-1/2}\phi_0(t)\,.\label{FT4}
\end{eqnarray}

\section{Phenomenological discussions}

With the unitary constraints, the new parametrization forms for   form factors of $B_c$ into the S-wave and P-wave charmonium in the entire $q^2$ range have been derived in the last section. The z-series expansion depends on the coefficients $\alpha_k$, which can be obtained by fitting the data or theoretical results. Although it contains an infinite tower of coefficients, one may expect that this expansion converges very fast, since the maximum of $z(t)$ is only 0.02 in the physical region $0\leq t\leq t_-$. Thus it is a good approximation to conserve the first order coefficient $\alpha_1$ and set $\alpha_i=0$ for $i>1$.

In the fitting, we will use  the lattice QCD data by the HPQCD Collaboration in Refs.~\cite{Colquhoun:2016osw,Lytle:2016ixw}   to determine  the form factors $f_+^{\eta_c}$, $f_0^{\eta_c}$, $V^{J/\psi}$,  and $A_1^{J/\psi}$. From the
fitting, one can see the LFQM predictions are  consistent with the HPQCD data. Besides, the lattice QCD results are also consistent with the HQET predictions at the minimal recoil point. Due to the lack of the lattice QCD data  for other form factors, we will employ the NRQCD and the HQET results to determine the rest form factors of  $B_c$ into the S-wave and P-wave charmonium.

\begin{figure}[t]
\centering
\includegraphics[width=0.48\linewidth]{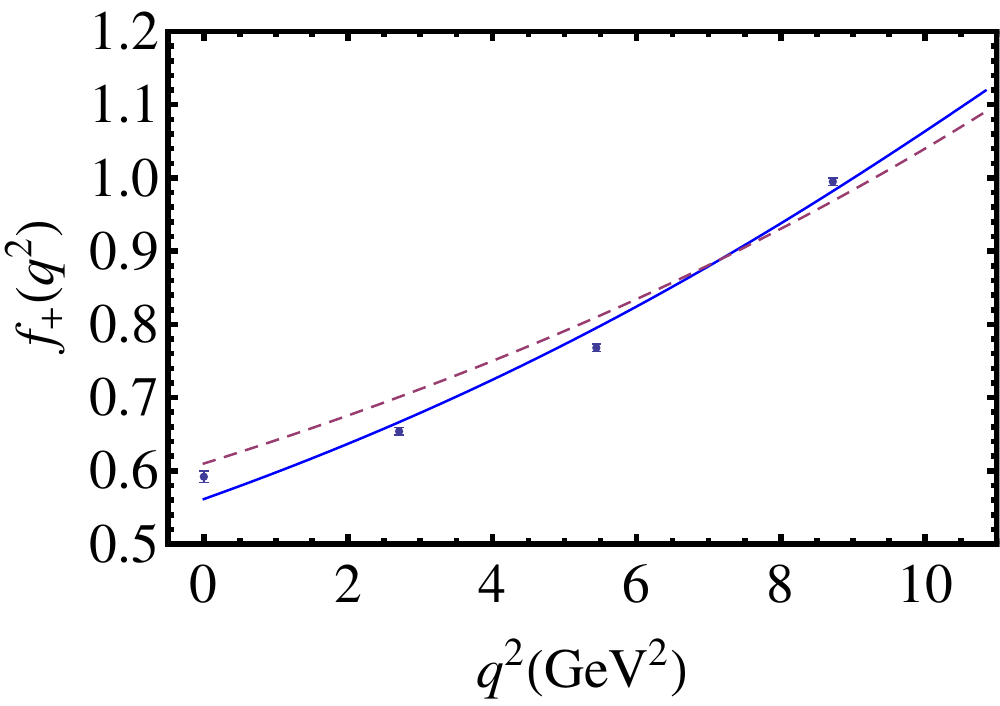}
\includegraphics[width=0.48\linewidth]{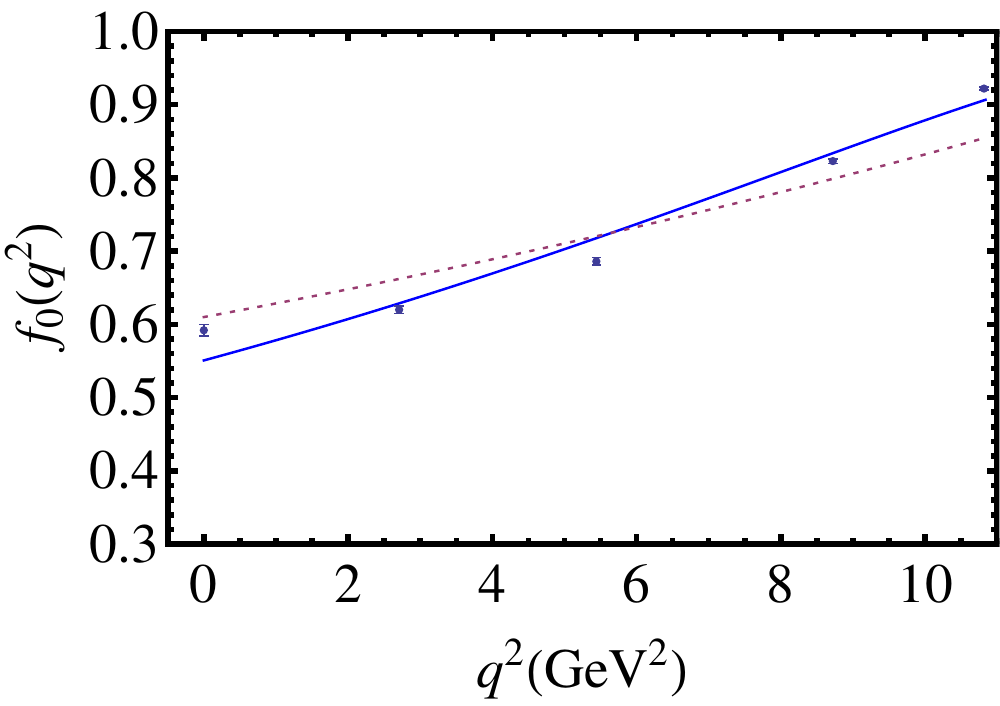}
\caption{The form factors of $B_c$ into $\eta_c$. The data is from the HPQCD lattice simulations~\cite{Colquhoun:2016osw}; the blue line is from the z-series based on the lattice data; the dashed line is from the LFQM results~\cite{Wang:2008xt}; the dotted line is from the the z-series based on the LFQM results.   }
\label{fig:etac}
\end{figure}

\begin{figure}[t]
\centering
\includegraphics[width=0.48\linewidth]{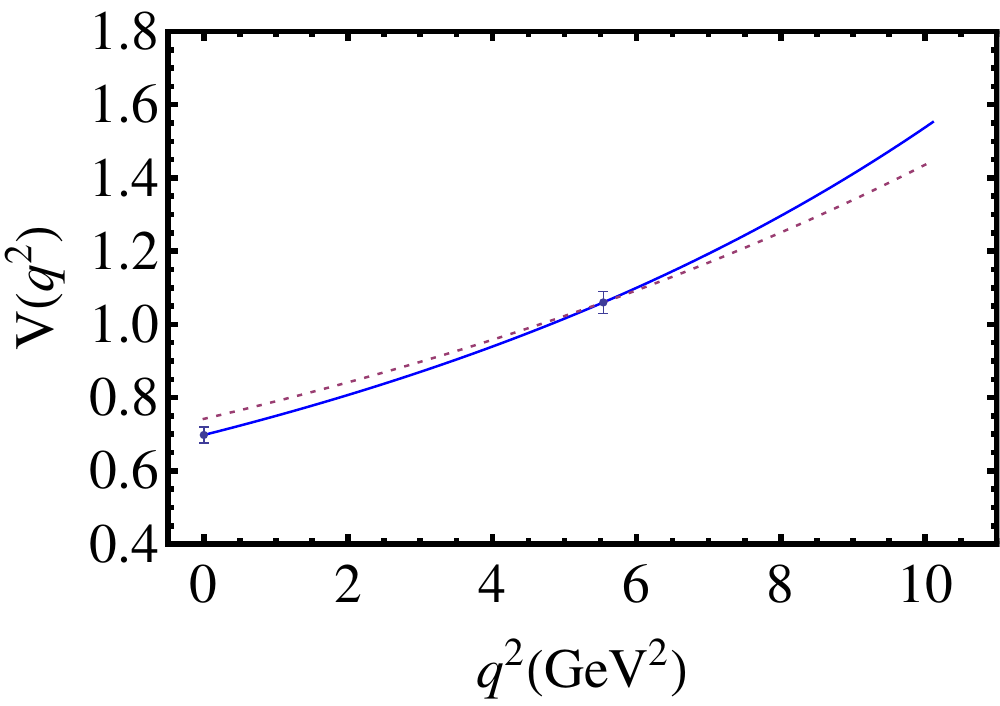}
\includegraphics[width=0.48\linewidth]{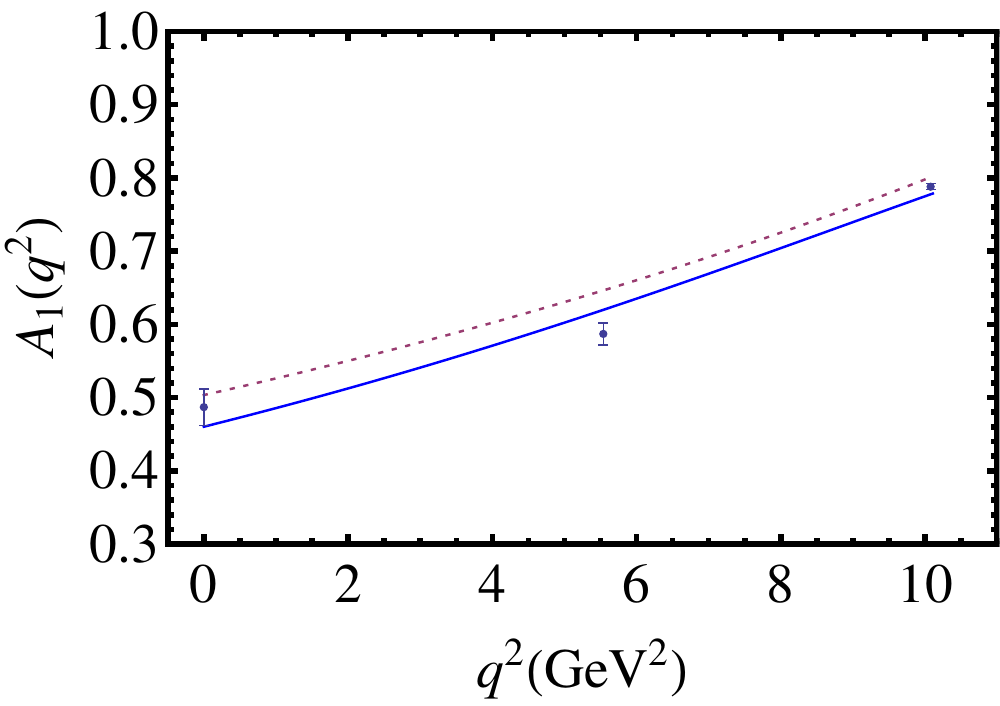}
\caption{The form factors of $B_c$ into $J/\psi$. The data is from the HPQCD lattice simulations~\cite{Colquhoun:2016osw}; the blue line is from the z-series based on the lattice data; the dashed line is from the LFQM results~\cite{Wang:2008xt}.}
\label{fig:jpsi1}
\end{figure}

\begin{figure}[t]
\centering
\includegraphics[width=0.48\linewidth]{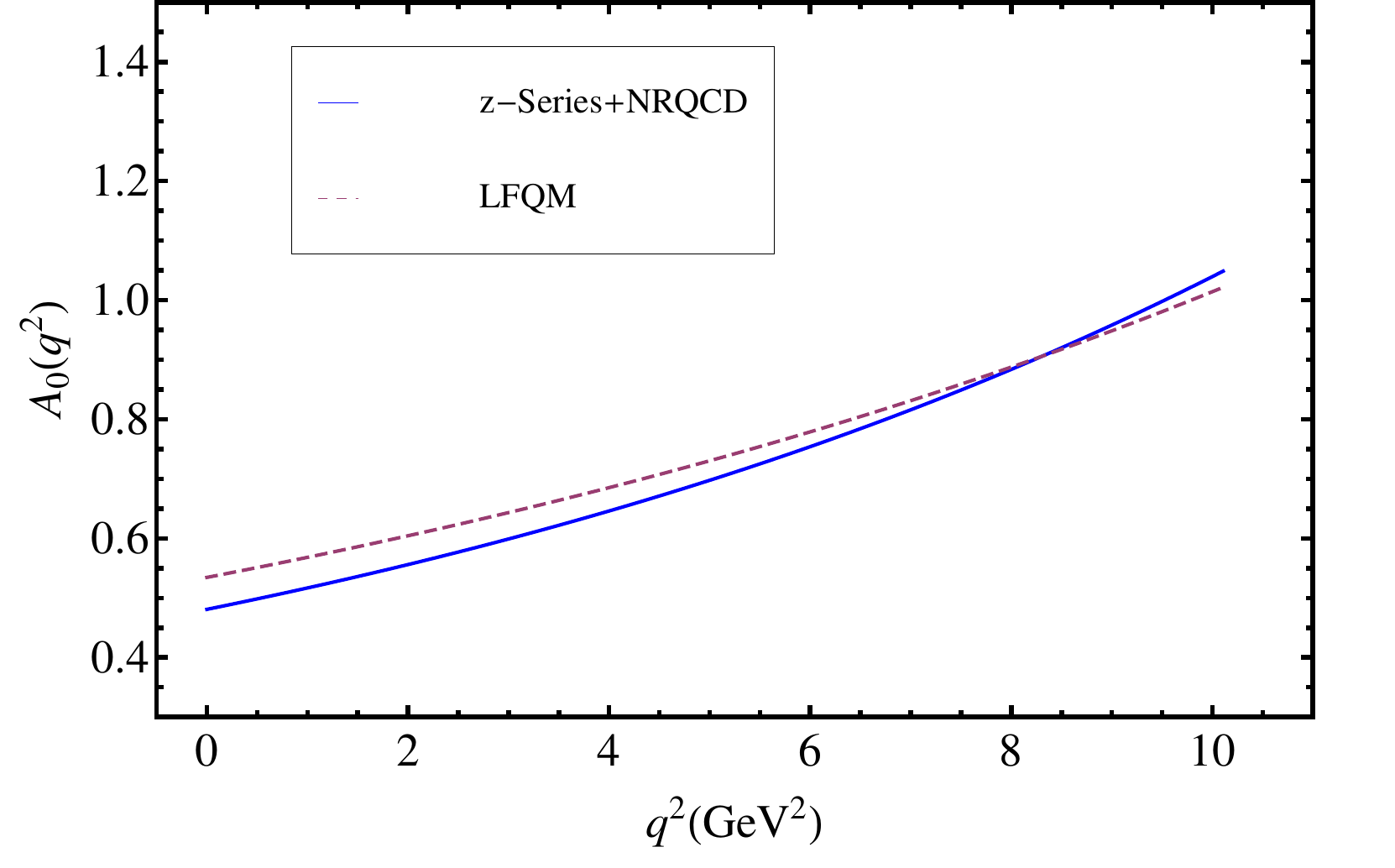}
\includegraphics[width=0.48\linewidth]{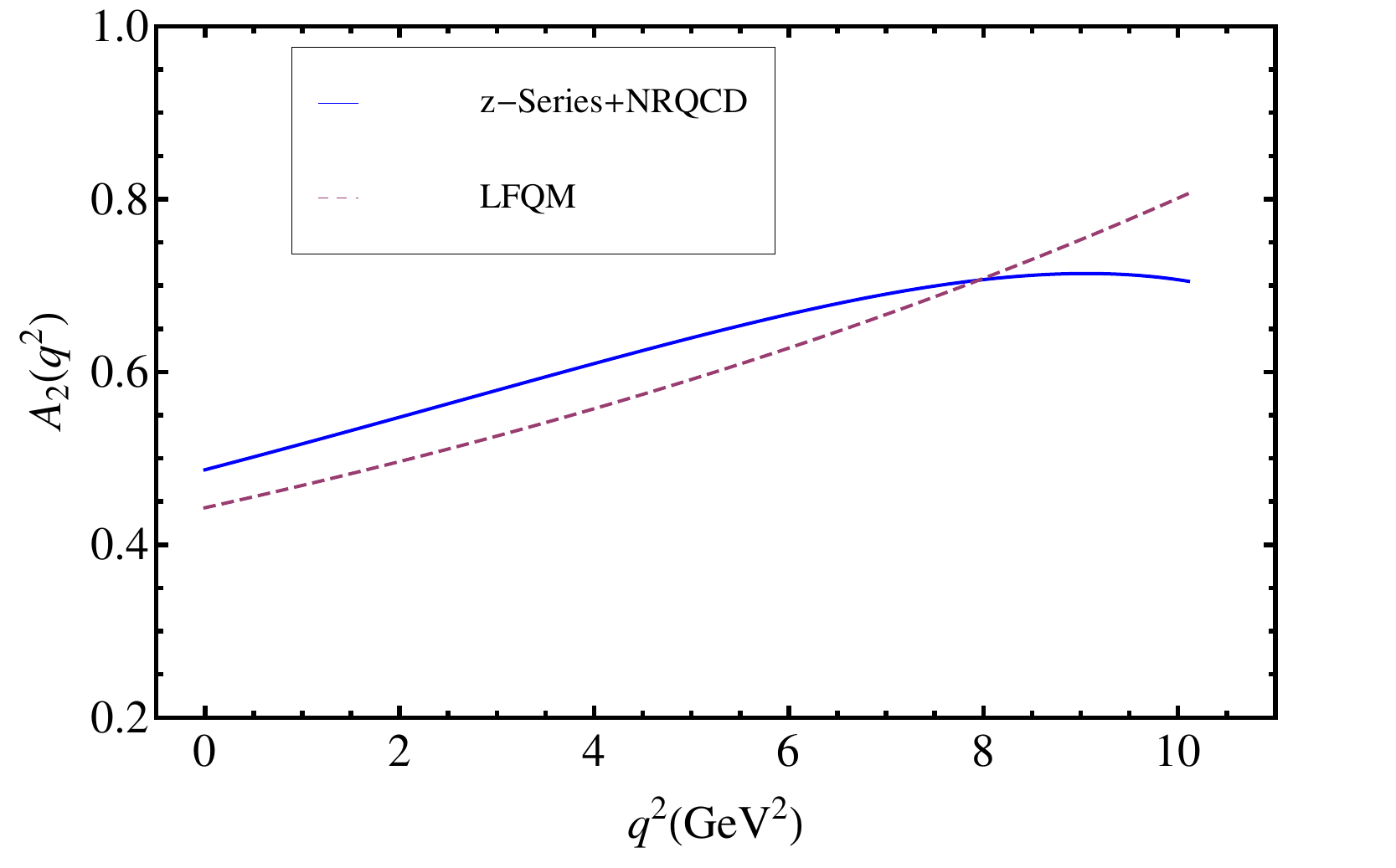}
\caption{The form factors of $B_c$ into $J/\psi$.  The dashed line is from the LFQM results~\cite{Wang:2008xt}; the blue line is from the the z-series based on the NRQCD results.  }
\label{fig:jpsi2}
\end{figure}

We have  shown  the form factors of $f_+^{\eta_c}$, $f_0^{\eta_c}$, $V^{J/\psi}$,  and $A_1^{J/\psi}$  in Figs.~\ref{fig:etac} and~\ref{fig:jpsi1}.
 and \ref{fig:jpsi2}. The heavy quark relations   for   form factors in Eqs.~(\ref{HQET1}-\ref{HQETn}) use   Isgur-Wise functions. At zero recoil with  $\omega=1$, the heavy quark limit gives  $f_+(\omega=1)=1.06$ and $f_0(\omega=1)=0.93$ for $B_c$ into $\eta_c$, and $V(\omega=1)=A_0(\omega=1)=A_2(\omega=1)=1.06$, $A_1(\omega=1)=0.94$ for $B_c$ into $J/\psi$. For $B_c$ into $\eta_c$ and $B_c$ into $J/\psi$, these predictions are consistent with  the z-series expansion.

\begin{figure}[t]
\centering
\includegraphics[width=0.48\linewidth]{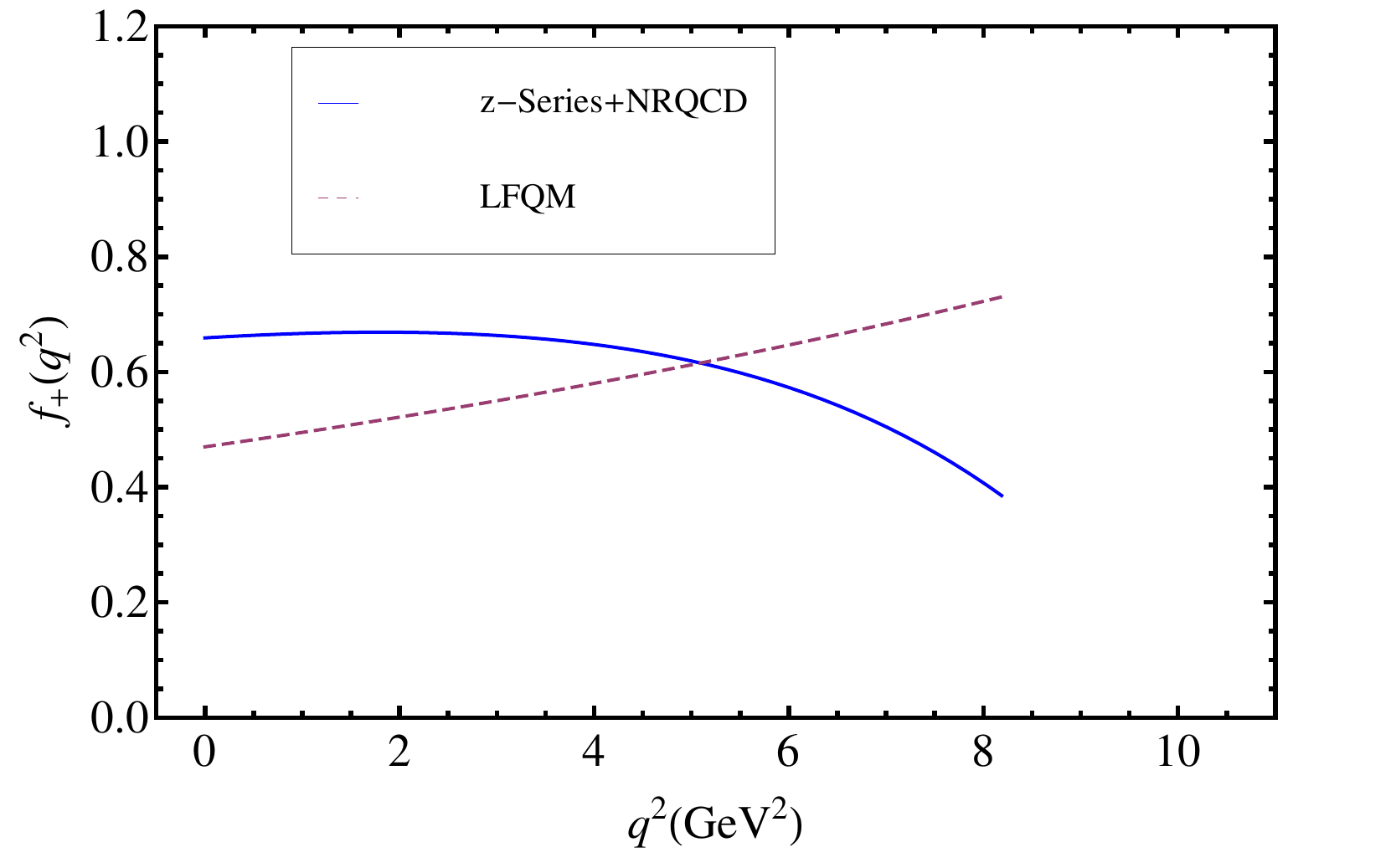}
\includegraphics[width=0.48\linewidth]{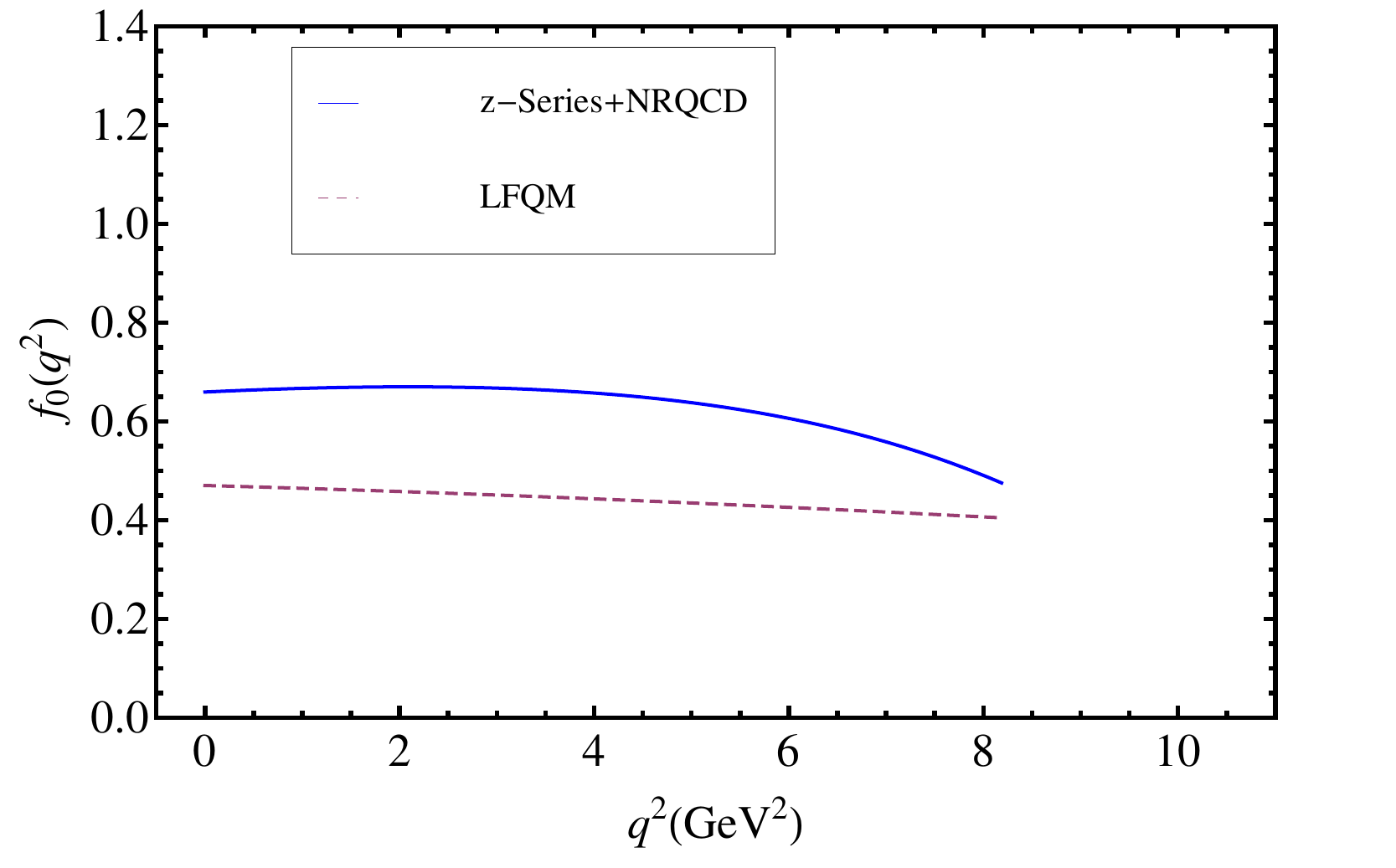}
\caption{The form factors of $B_c$ into $\chi_{c0}$.  The dashed line is from the LFQM results~\cite{Wang:2009mi}; the blue line is from the the z-series based on the NRQCD results.  }
\label{fig:chic0}
\end{figure}

\begin{figure}[t]
\centering
\includegraphics[width=0.48\linewidth]{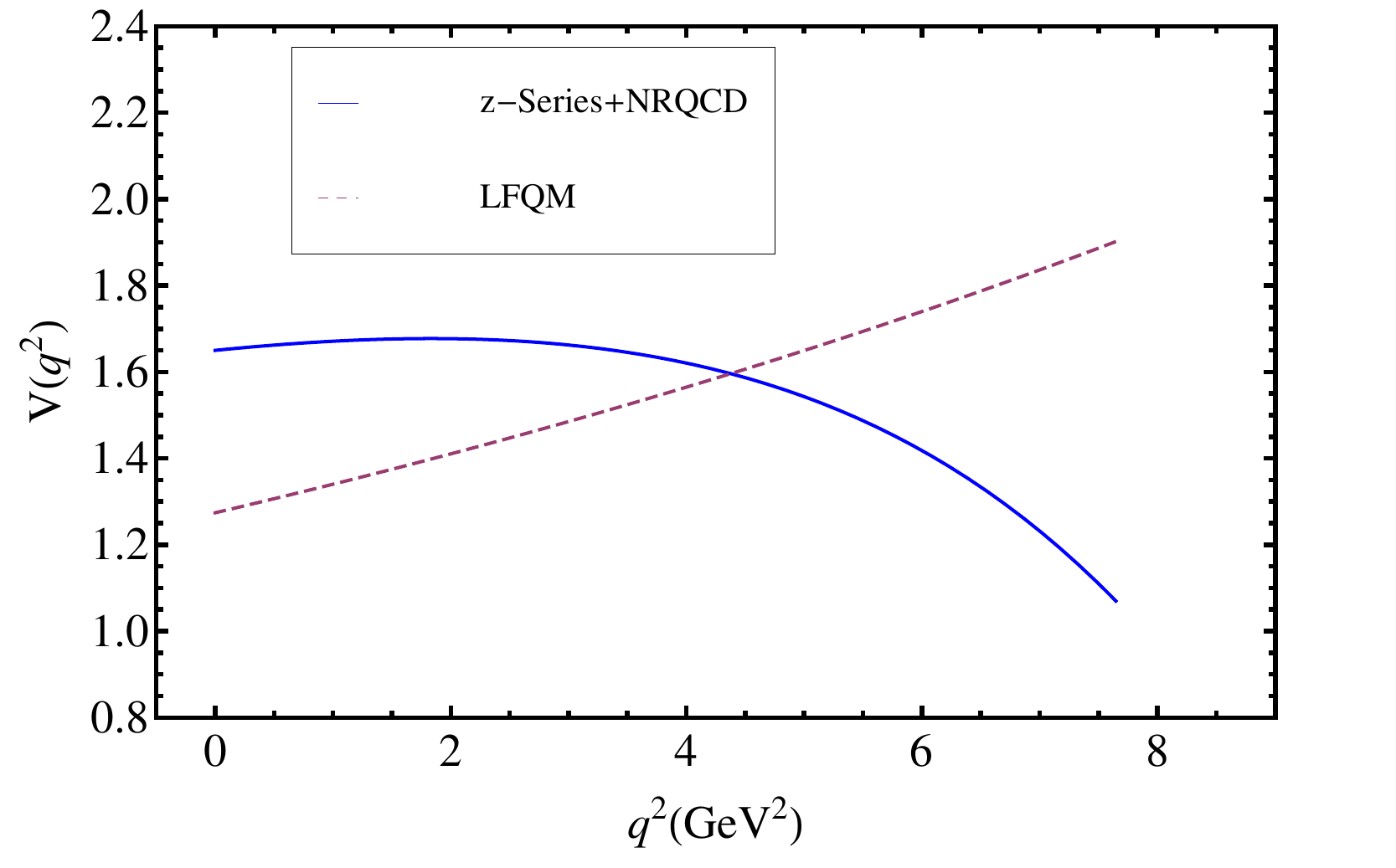}
\includegraphics[width=0.48\linewidth]{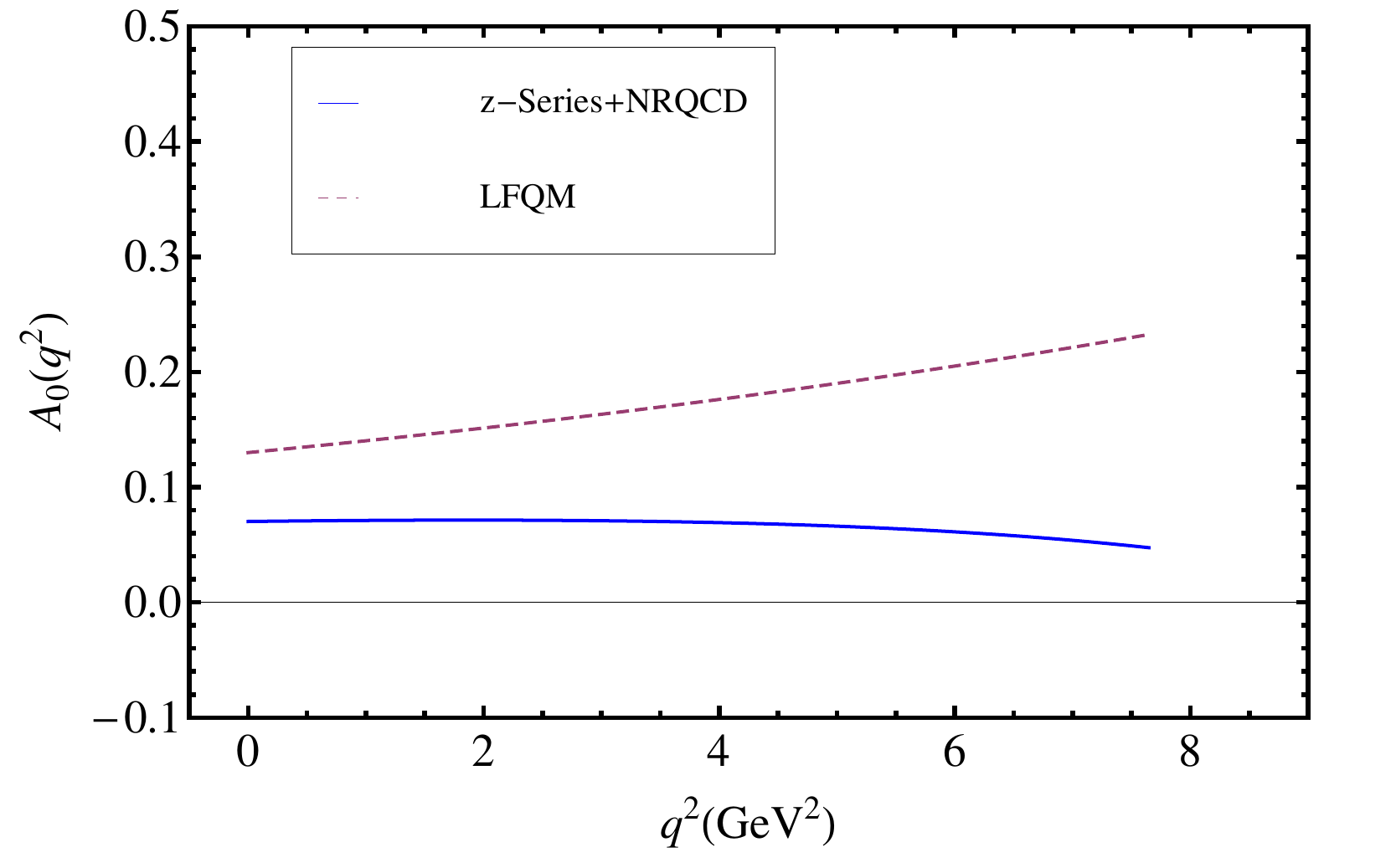}
\includegraphics[width=0.48\linewidth]{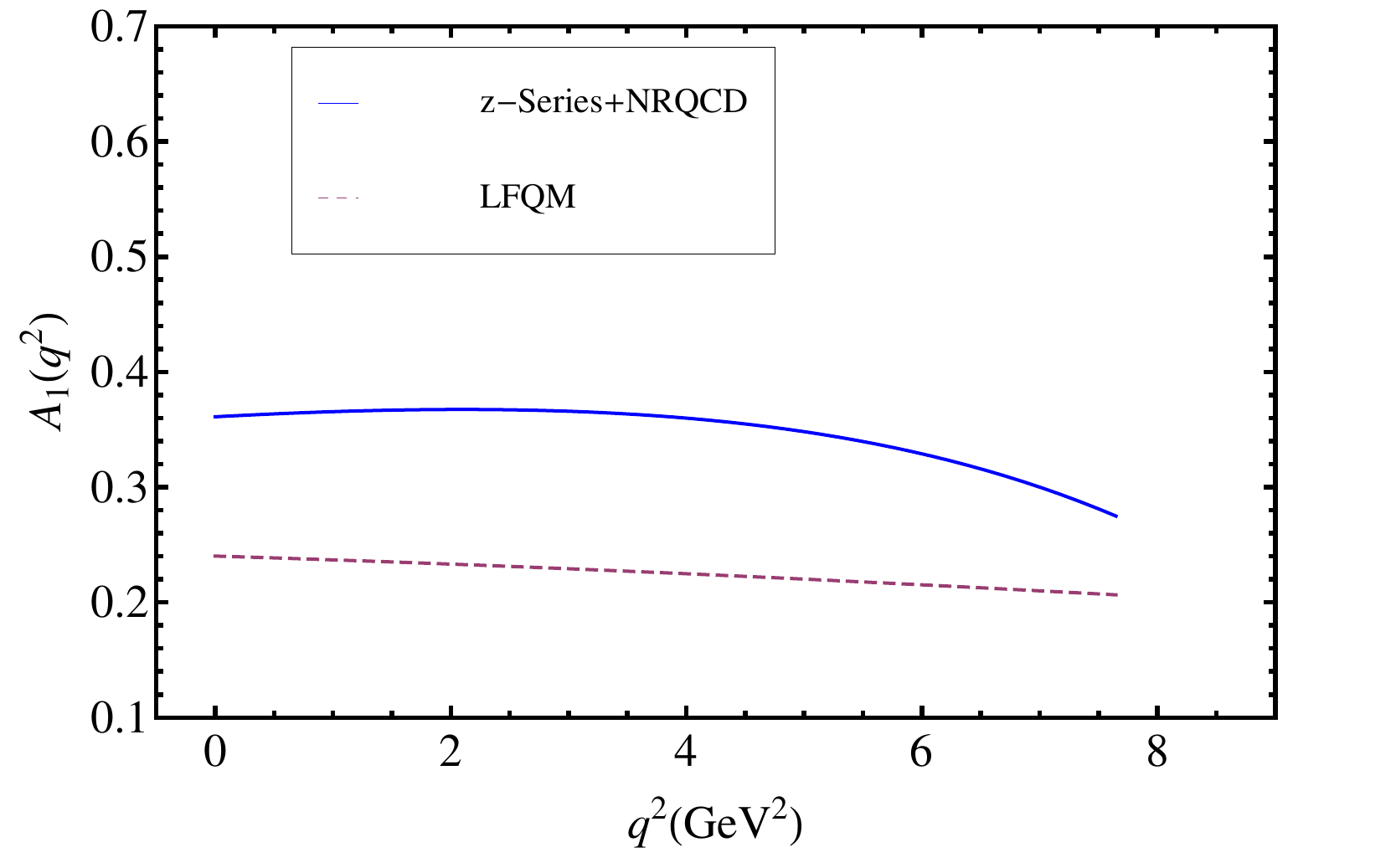}
\includegraphics[width=0.48\linewidth]{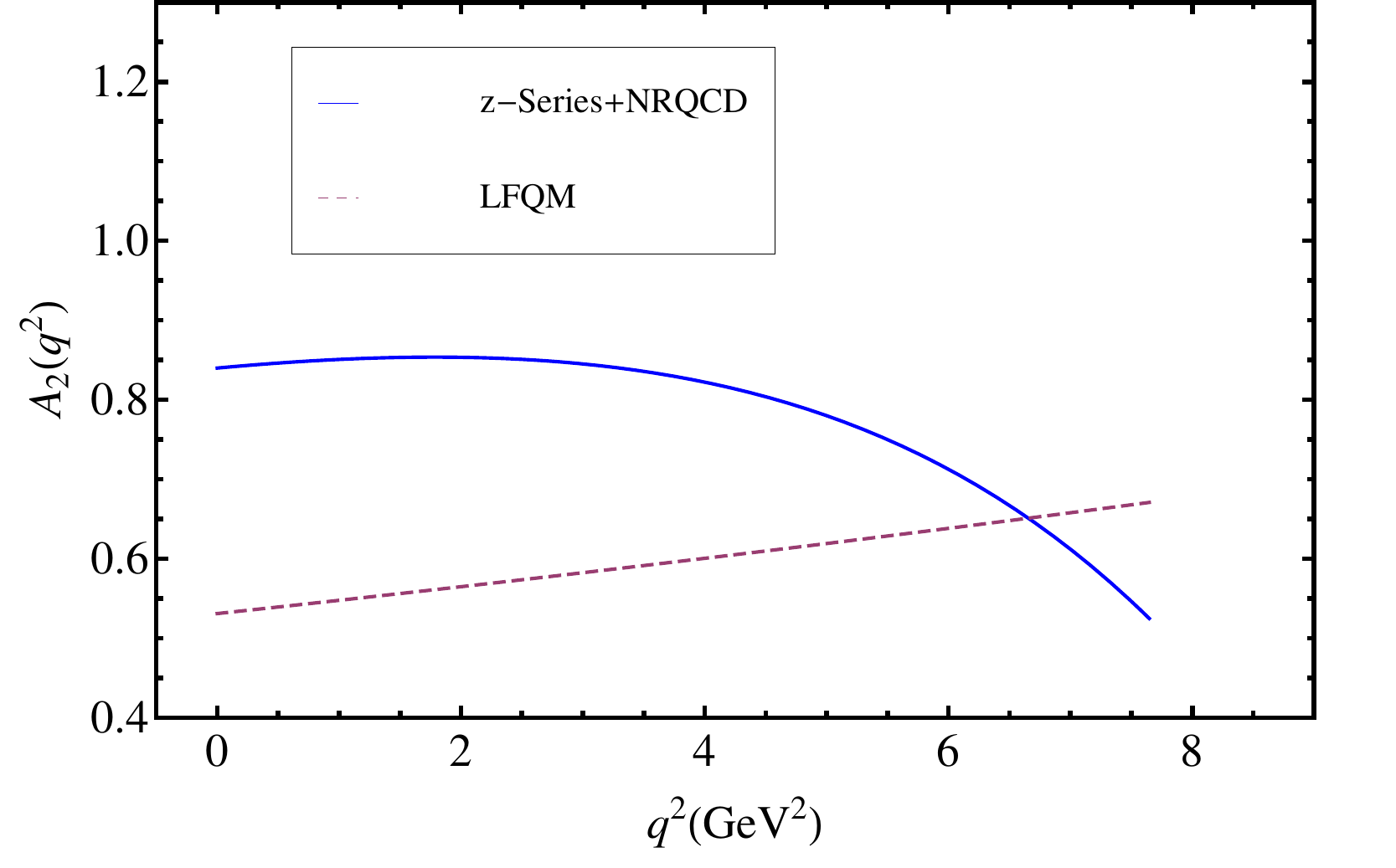}
\caption{The form factors of $B_c$ into $\chi_{c1}$.  The dashed line is from the LFQM results~\cite{Wang:2009mi}; the blue line is from the the z-series based on the NRQCD results.  }
\label{fig:chic1}
\end{figure}

\begin{figure}[t]
\centering
\includegraphics[width=0.48\linewidth]{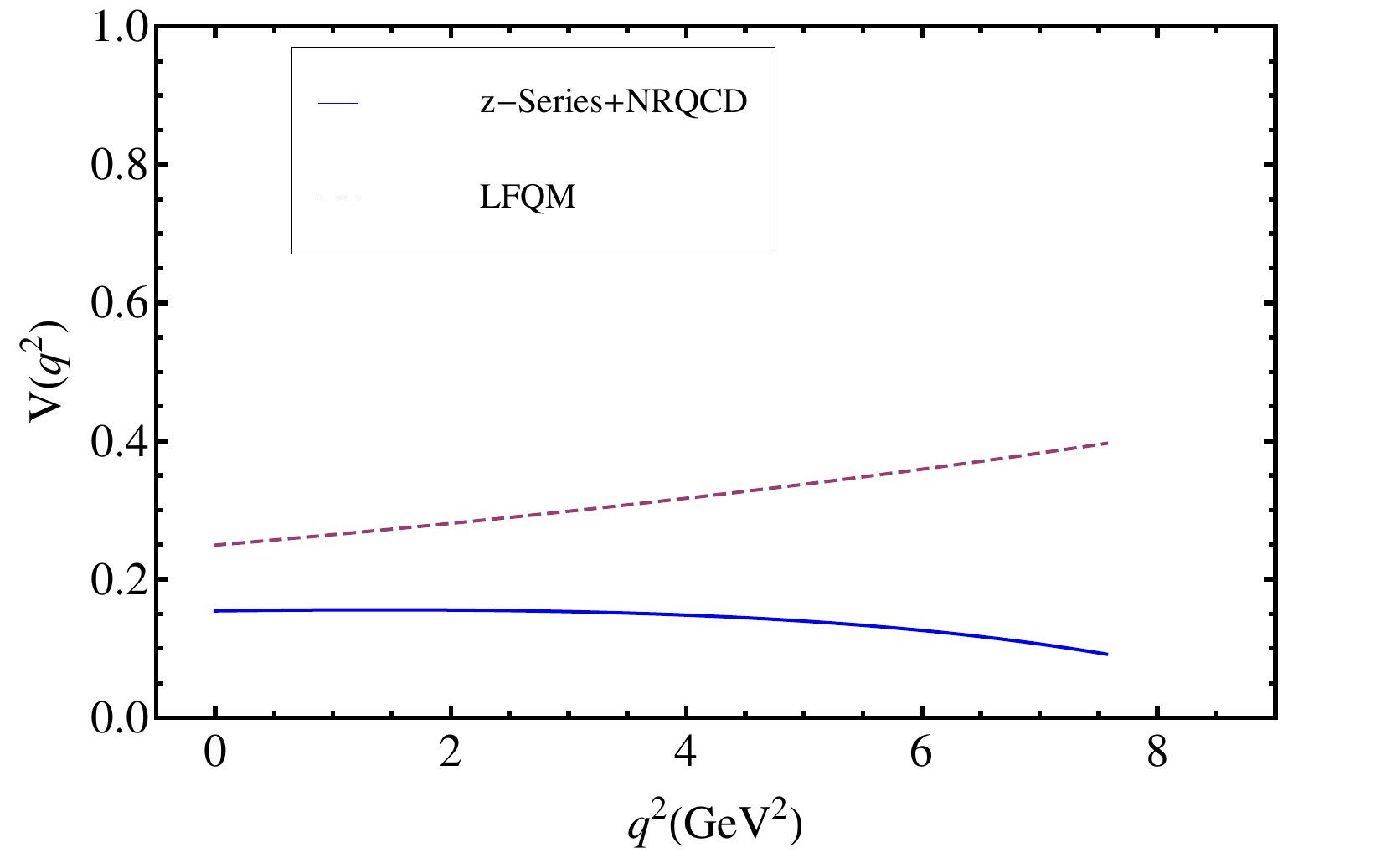}
\includegraphics[width=0.48\linewidth]{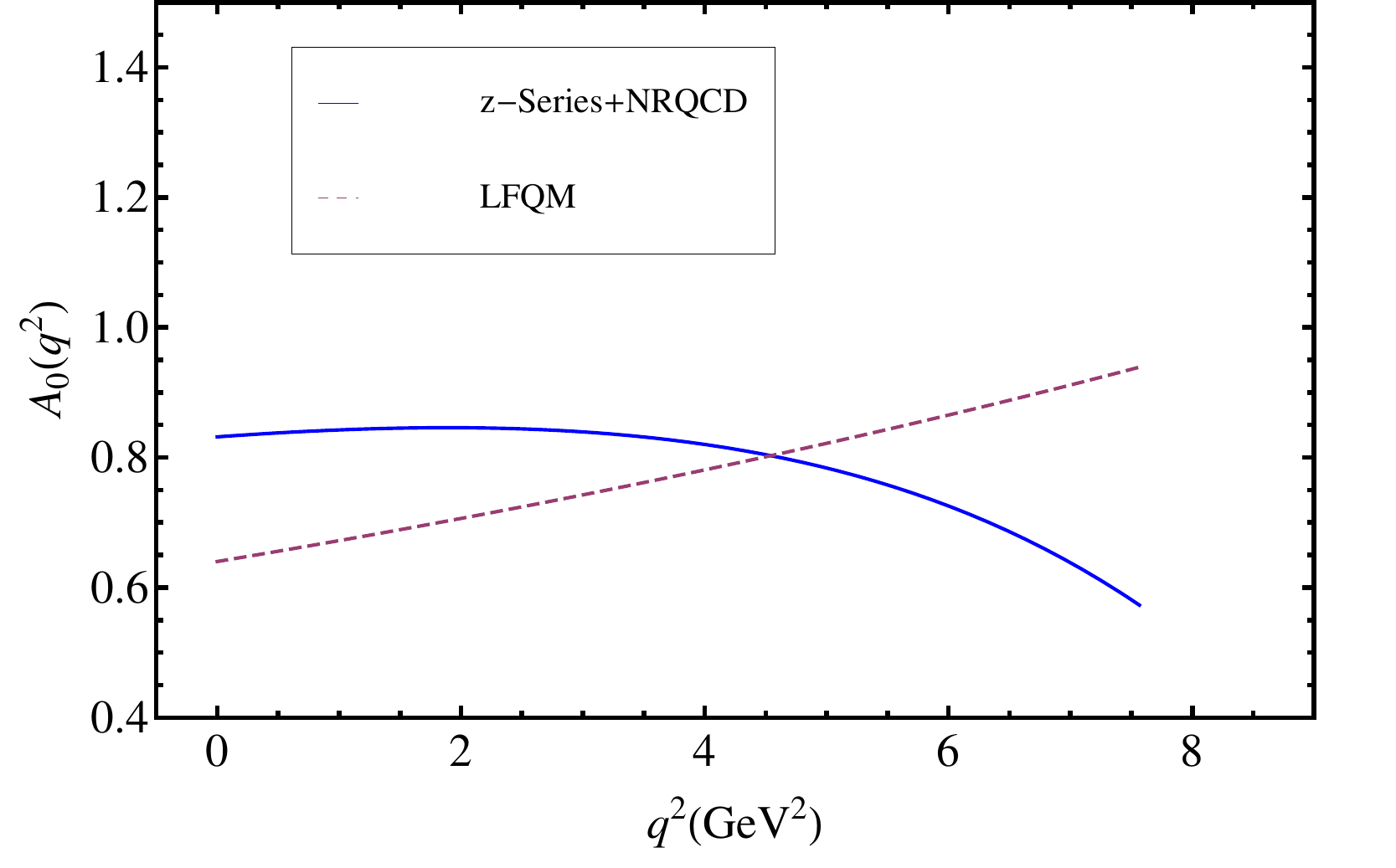}
\includegraphics[width=0.48\linewidth]{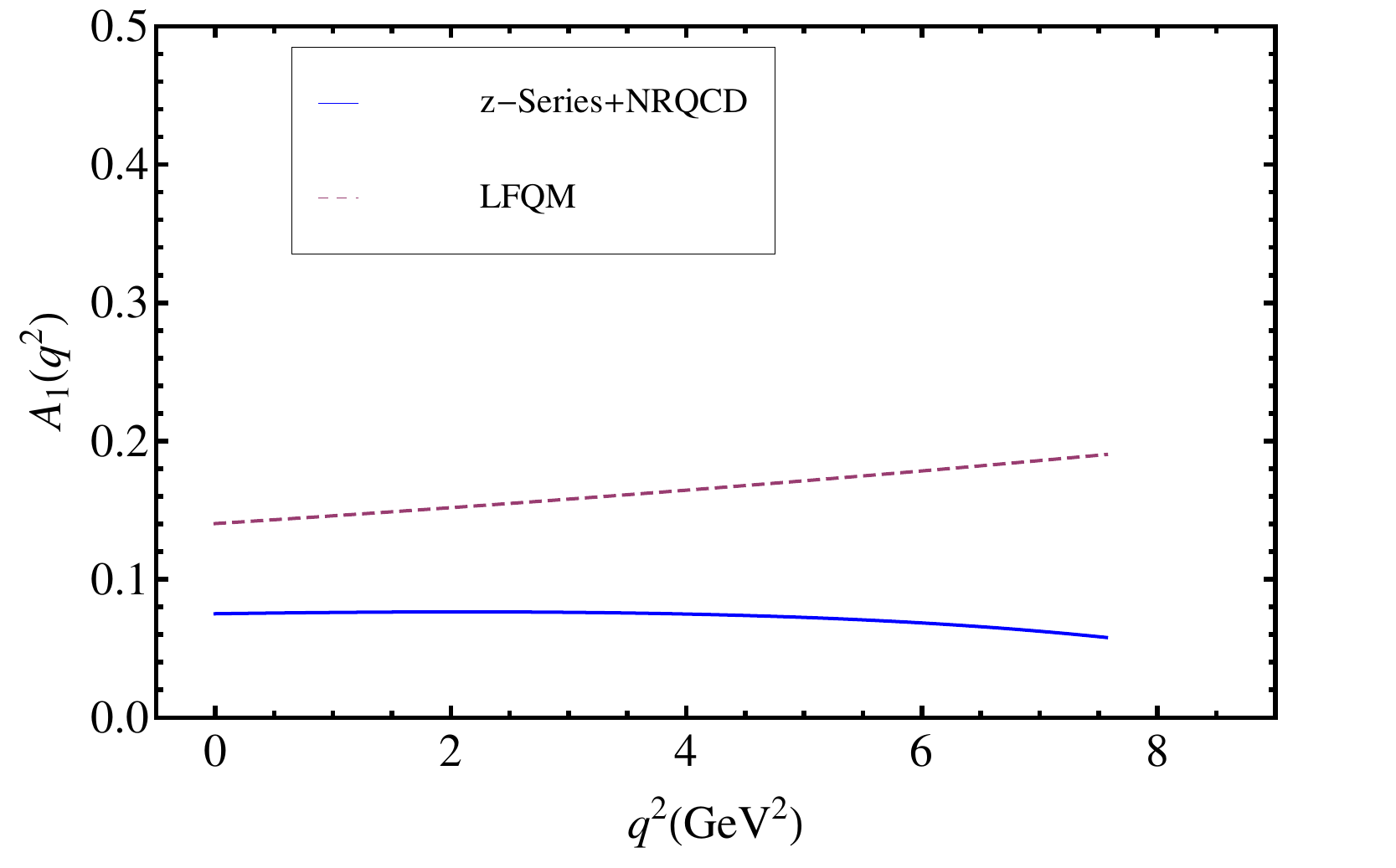}
\includegraphics[width=0.48\linewidth]{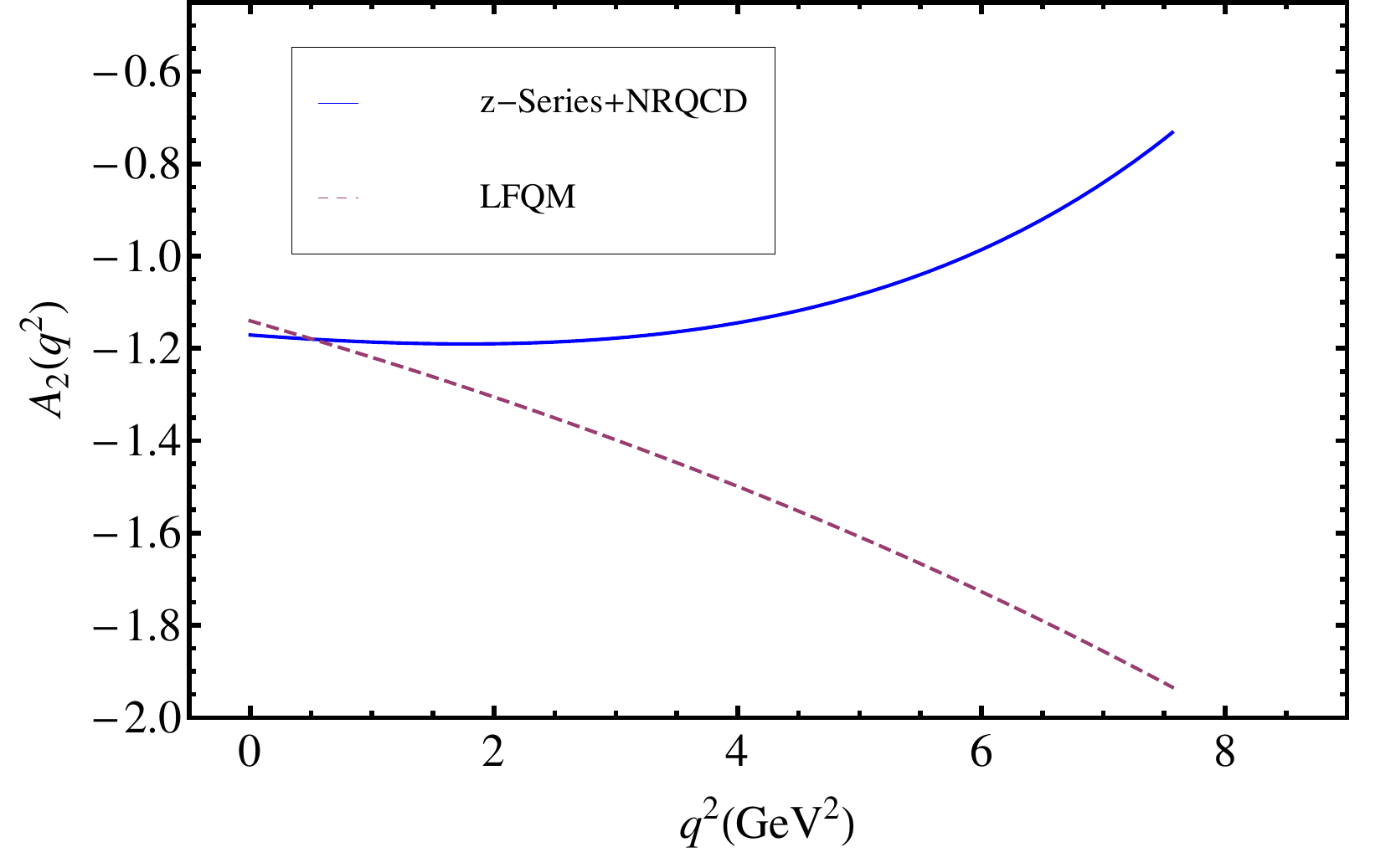}
\caption{The form factors of $B_c$ into $h_{c}$.  The dashed line is from the LFQM results~\cite{Wang:2009mi}; the blue line is from the the z-series based on the NRQCD results.  }
\label{fig:hc}
\end{figure}

\begin{figure}[t]
\centering
\includegraphics[width=0.48\linewidth]{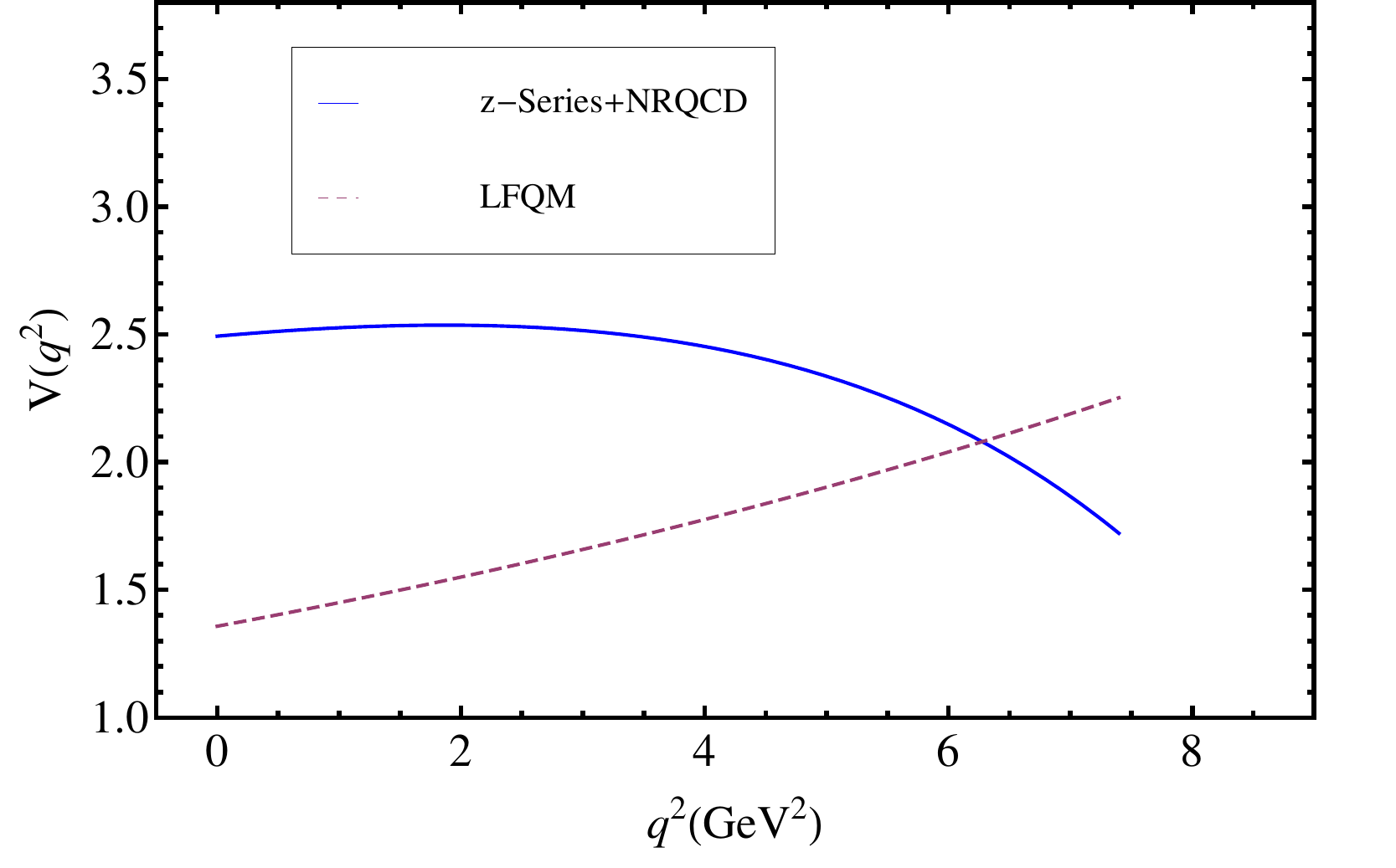}
\includegraphics[width=0.48\linewidth]{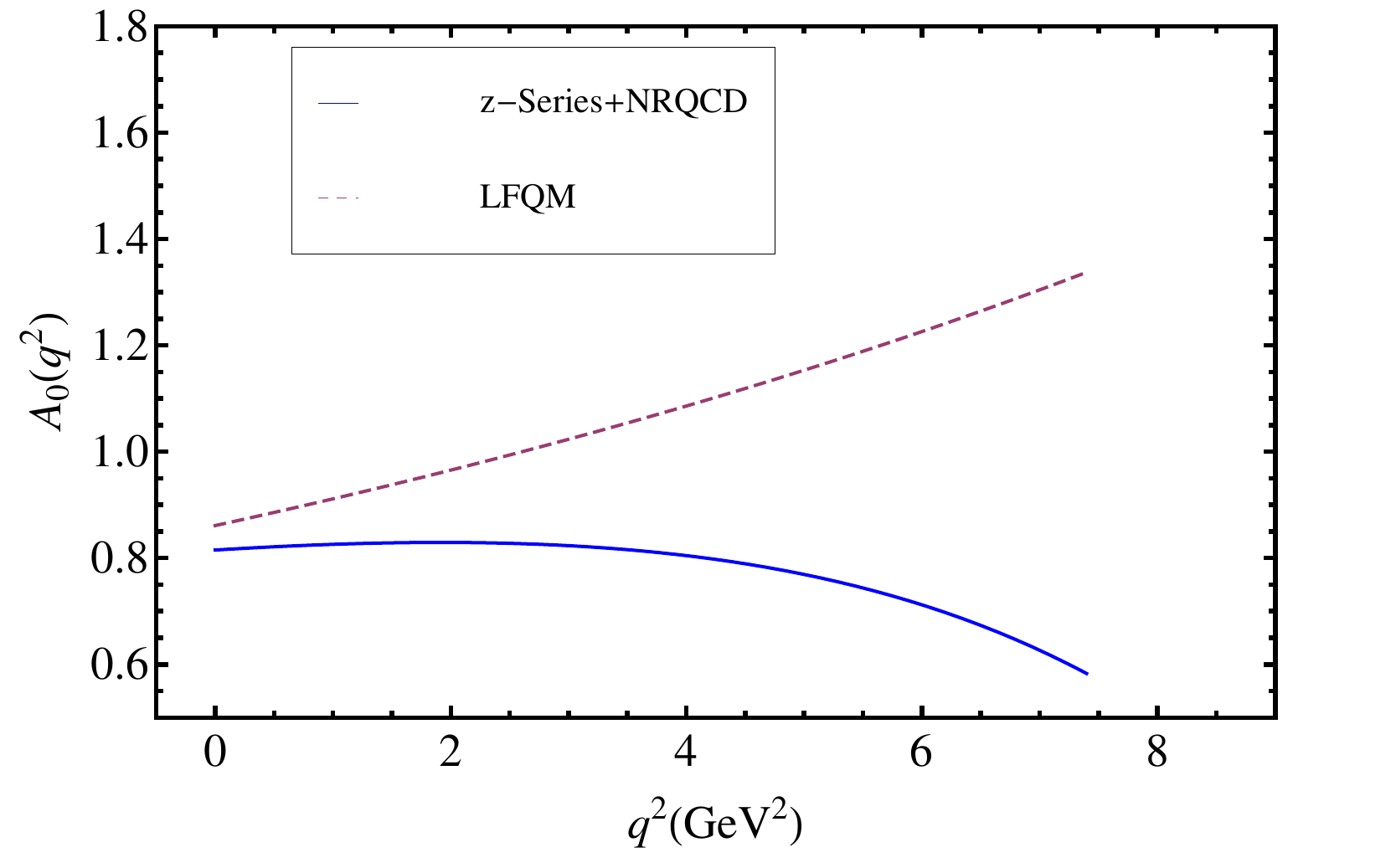}
\includegraphics[width=0.48\linewidth]{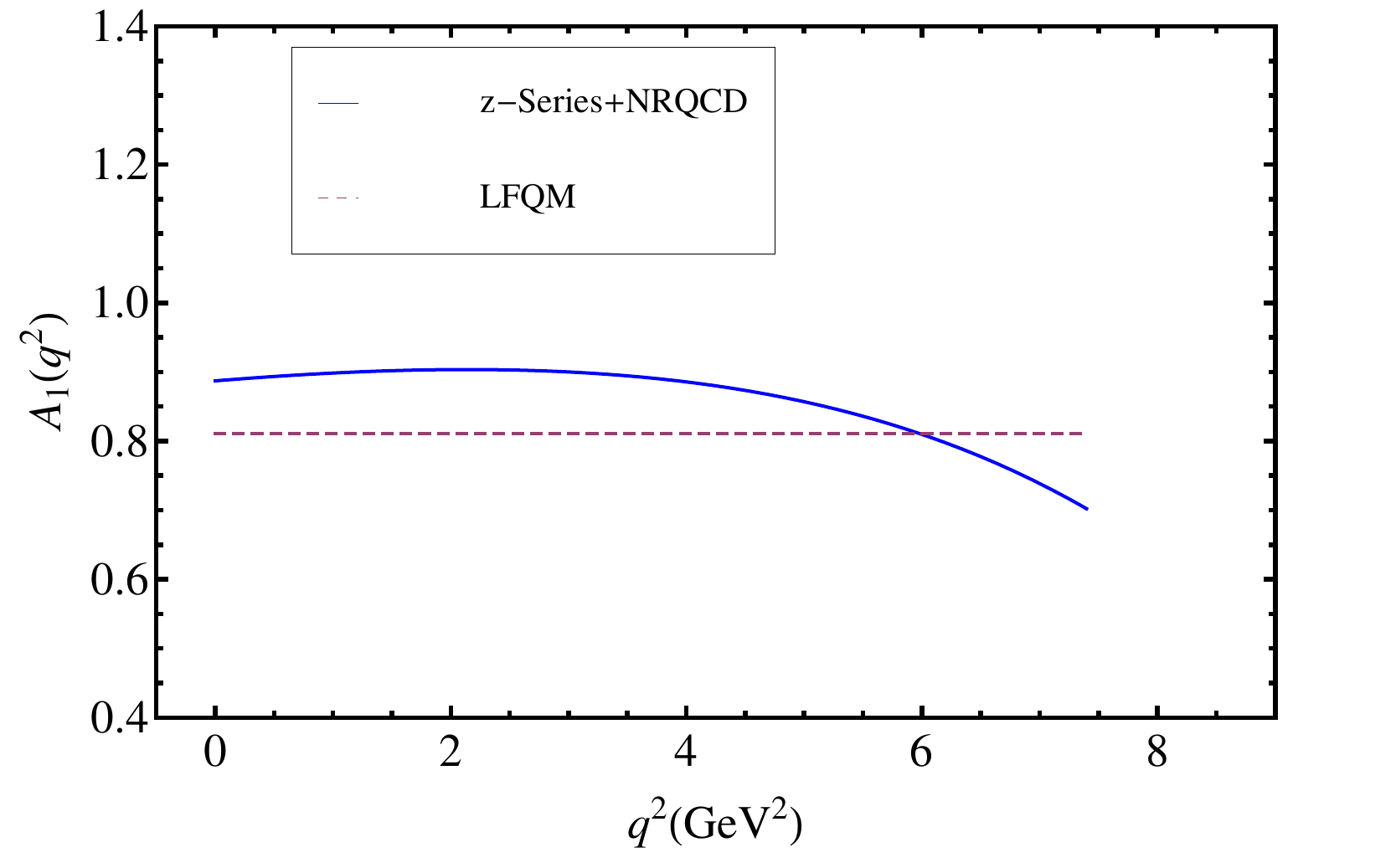}
\includegraphics[width=0.48\linewidth]{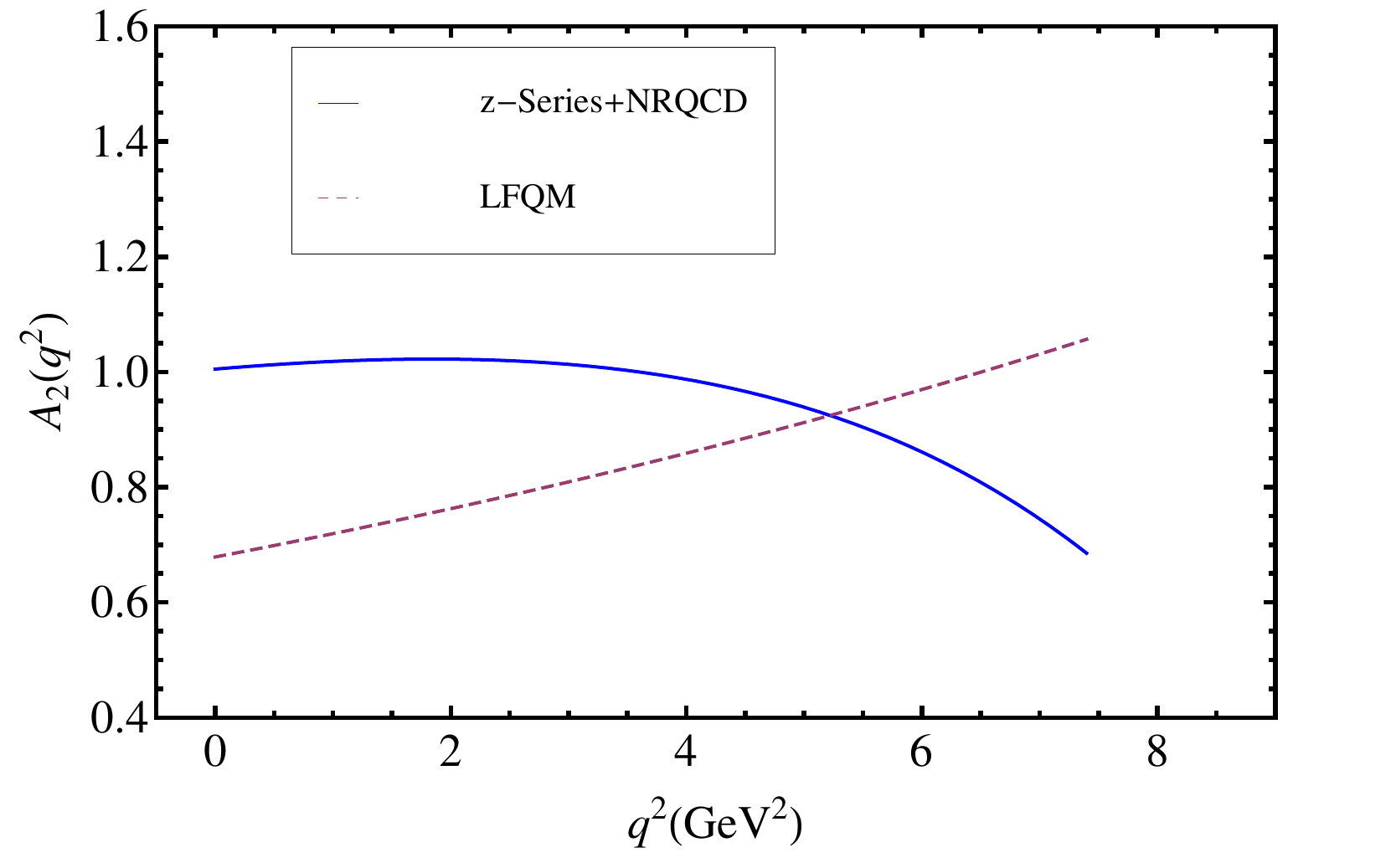}
\caption{The form factors of $B_c$ into $\chi_{c2}$.  The dashed line is from the LFQM results~\cite{Wang:2009mi}; the blue line is from the the z-series based on the NRQCD results.  }
\label{fig:chic2}
\end{figure}

We have also plotted the form factors of $B_c$ into P-wave charmonium  within z-series method in Figs.~\ref{fig:chic0}, \ref{fig:chic1}, \ref{fig:hc} and \ref{fig:chic2}. These results are helpful to perform the phenomenological analysis.

Semileptonic decays  $B_c\to \eta_c \ell\bar\nu_{\ell}$ have the decay widths:
\begin{eqnarray}
 \frac{d\Gamma(B_c\to Pl\bar\nu_l)}{dq^2} &=&(\frac{q^2-m_l^2}{q^2})^2\frac{ {\sqrt{\lambda(m_{B_c}^2,m_P^2,q^2)}} G_F^2 |V_{\rm CKM}|^2} {384m_{B_c}^3\pi^3}
  \frac{1}{q^2} \nonumber\\
&\times& \left\{ (m_l^2+2q^2) \lambda(m_{B_c}^2,m_P^2,q^2) f_+^2(q^2)  +3 m_l^2(m_{B_c}^2-m_P^2)^2f_0^2(q^2)
 \right\},
 \end{eqnarray}
where $V_{\rm CKM}=V_{cb}$ and  $\lambda(m_{B_c}^2,
m_{i}^2,q^2)=(m_{B_c}^2+m_{i}^2-q^2)^2-4m_{B_c}^2m_i^2$. This formula is also valid for semileptonic decays of $B_c$ into a scalar charmonium by $P\to S$.

Decay widths for $B_c\to J/\psi\ell\bar\nu_{\ell}$ are given as:
 \begin{eqnarray}
 \frac{d\Gamma_L(B_c\to Vl\bar\nu)}{dq^2}&=&(\frac{q^2-m_l^2}{q^2})^2\frac{ {\sqrt{\lambda(m_{B_c}^2,m_V^2,q^2)}} G_F^2 |V_{\rm CKM}|^2} {384m_{B_c}^3\pi^3}
 \frac{1}{q^2} \nonumber\\
 &\times& \left\{ 3 m_l^2 \lambda(m_{B_c}^2,m_V^2,q^2) A_0^2(q^2)+(m_l^2+2q^2) |\frac{1}{2m_V}  \left[
 (m_{B_c}^2-m_V^2-q^2)\right.\right.\nonumber\\
 &\times&\left.\left.
 (m_{B_c}+m_V)A_1(q^2)-\frac{\lambda(m_{B_c}^2,m_V^2,q^2)}{m_{B_c}+m_V}A_2(q^2)\right]|^2
 \right\},\label{eq:decaywidthlon}\\
 \frac{d\Gamma^\pm(B_c\to
 Vl\bar\nu)}{dq^2}&=&(\frac{q^2-m_l^2}{q^2})^2\frac{ {\sqrt{\lambda(m_{B_c}^2,m_V^2,q^2)}} G_F^2 |V_{\rm CKM}|^2} {384m_{B_c}^3\pi^3}
    \nonumber\\
 &\times& \left\{ (m_l^2+2q^2) \lambda(m_{B_c}^2,m_V^2,q^2)\left|\frac{V(q^2)}{m_{B_c}+m_V}\mp
 \frac{(m_{B_c}+m_V)A_1(q^2)}{\sqrt{\lambda(m_{B_c}^2,m_V^2,q^2)}}\right|^2
 \right\},\;
\end{eqnarray}
where we use the subscript $+(-)$ to denote the right-handed (left-handed)
states of vector mesons. The total and transverse differential decay widths
are given by:
\begin{eqnarray}
 \frac{d\Gamma}{dq^2}= \frac{d\Gamma_L}{dq^2}+
 \frac{d\Gamma_\perp}{dq^2},\;\;\;
\frac{d\Gamma_\perp}{dq^2}= \frac{d\Gamma^+}{dq^2}+
 \frac{d\Gamma^-}{dq^2}.
\end{eqnarray}
The formulae are also valid for decays of $B_c$ into axial-vector charmonia by $V\to \chi_{c1}(h_c)$.

For the tensor meson in the final state, the semileptonic $B_c$ decays have
\begin{eqnarray}
 \frac{d\Gamma_L(B_c\to T l\bar{\nu}_l)}{dq^2}&=& \frac{2}{3}\frac{ \lambda(m_{B_c}^2,m_T^2,q^2)}{4m_T^2m_{B_c}^2}\frac{d\Gamma_L(B_c\to V l\bar{\nu}_l)}{dq^2}|_{A_i^V\to A_i^T},\nonumber\\
\frac{d\Gamma^{\pm}(B_c\to T l\bar{\nu}_l)}{dq^2}&=& \frac{1}{2}\frac{ \lambda(m_{B_c}^2,m_T^2,q^2)}{4m_T^2m_{B_c}^2}\frac{d\Gamma^{\pm}(B_c\to V l\bar{\nu}_l)}{dq^2}|_{A_i^V\to A_i^T,V^V\to V^T}.
\end{eqnarray}

In the numerical calculation, we adopt the inputs as: $m_{B_c}=6.276$GeV, $m_{\eta_c}=2.984$GeV, $m_{J/\psi}=3.097$GeV, $m_{h_c}=3.525$GeV, $m_{\chi_{c0}}=3.415$GeV, $m_{\chi_{c1}}=3.511$GeV, $m_{\chi_{c2}}=3.556$GeV, $\tau_{B_c}=0.503$ps~\cite{Agashe:2014kda}.  We adopt the heavy quark masses  as $m_c=(1.4\pm0.2)$GeV and $m_b=(4.8\pm 0.2)$GeV~\cite{Wang:2015bka,Zhu:2016udl,Zhu:2015bba,Wang:2017vnc}. We find  that the ratios between $B_c\to H+\mu +\bar\nu_\mu $ and $B_c\to H+e^- +\bar\nu_e $
is rather close to 1, which is under expectation. For the semitaunic and semimuonic $B_c$ decays into a charmonium $H$, the $R_H$ and the helicity dependent ratios are defined as
\begin{eqnarray}
 R_H&=& \frac{\Gamma(B_c\to H+\tau +\bar\nu_\tau)}{\Gamma(B_c\to H+\mu +\bar\nu_\mu)},\\
 R^L_H&=& \frac{\Gamma_L(B_c\to H+\tau +\bar\nu_\tau)}{\Gamma_L(B_c\to H+\mu +\bar\nu_\mu)},\\
 R^\perp_H&=& \frac{\Gamma_\perp(B_c\to H+\tau +\bar\nu_\tau)}{\Gamma_\perp(B_c\to H+\mu +\bar\nu_\mu)}.
\end{eqnarray}

We give the branching ratios of $B_c\to H+\mu +\bar\nu_\mu $ in Tab.~\ref{tab:results1}. The ratios of the semitaunic and semimuonic $B_c$ decays into a charmonium $H$ are  predicted in  Tab.~\ref{tab:results2}. From this table, we can see the uncertainties are reduced for the $R_H$ since most uncertainties in form factors cancel. The SM prediction  for $R_{J/\psi}$ is far below the LHCb data. If confirmed, this  may indicate the possible new physics.  In this case, these NP effects  should also manifest themselves in the $B_c$ decays into P-wave charmonium, where we found the SM results for the ratios are even smaller.
Enhancements of these ratios on the experimental side can reveal the presence of NP further.

\begin{table}[thb]
\caption{\label{tab:results1}
Branching ratios (\%) of $B_c\to H+\mu +\bar\nu_\mu $ using  the z-series expanded  and LFQM form factors.   }
\begin{center}
\begin{tabular}{cccc}
  \hline \hline
  Channels & LFQM~\cite{Wang:2008xt,Wang:2009mi} & z-series+NRQCD \\  \hline
 $ B_c \to \eta_c \mu \bar\nu_\mu$  & $0.67\pm0.10$  & $0.66\pm0.02$ \\
$ B_c \to J/\psi \mu \bar\nu_\mu$   & $1.49\pm0.27$& $1.44\pm0.02$ \\
  $ B_c \to \chi_{c0} \mu \bar\nu_\mu$ &  $0.21\pm0.04$ & $0.33^{+0.03}_{-0.02}$ \\
 $ B_c \to \chi_{c1} \mu \bar\nu_\mu$ & $0.14\pm0.02$ &$0.11\pm0.03$ \\
 $ B_c \to h_c \mu \bar\nu_\mu$  & $0.31\pm0.08$  & $0.17\pm0.02$\\
  $ B_c \to \chi_{c2} \mu \bar\nu_\mu$ & $0.17\pm0.06$  & $0.17\pm0.04$  \\
   \hline \hline
\end{tabular}
\end{center}
\end{table}

\begin{table}[thb]
\caption{\label{tab:results2}
Predictions for the $R_H$ in  the z-series expansion  approach, where the uncertainty in the z-Series approach is from the form factors, and the
the uncertainty in the new physics predictions  is from the new Wilson coefficients. }
\begin{center}
\begin{tabular}{ccccc}
  \hline \hline
 $R_H$  & LHCb data~\cite{Aaij:2017tyk} ~~& z-Series+NRQCD& S1 &S2\\  \hline
$R_{J/\psi}$  &$0.71\pm0.17(stat)\pm0.18(syst)$ & $0.26\pm0.01$ & $0.32\pm0.02$ & $0.32\pm0.01$ \\
$R^L_{J/\psi}$  & & $0.24\pm0.01$ &$0.29\pm0.01$ & $0.28\pm0.01$\\
  $R^\perp_{J/\psi}$  & & $0.29\pm0.01$ &$0.36\pm0.02$ & $0.35\pm0.01$\\
 $R_{\eta_c}$  & & $0.31\pm0.01$ & $0.39\pm0.02$ & $0.41\pm0.02$ \\
 $R_{\chi_{c0}}$  & & $0.11\pm0.01$ & $0.11\pm0.01$ & $0.12\pm0.01$ \\
$R_{\chi_{c1}}$  & & $0.10\pm0.01$ & $0.11\pm0.01$ & $0.12\pm0.01$ \\
$R^L_{\chi_{c1}}$  & & $0.10\pm0.01$ & $0.11\pm0.02$ & $0.13\pm0.01$\\
  $R^\perp_{\chi_{c1}}$  & & $0.10\pm0.01$ & $0.11\pm0.01$ & $0.12\pm0.01$\\
  $R_{h_c}$  & & $0.06^{+0.03}_{-0.01}$ &$0.08^{+0.03}_{-0.01}$& $0.08^{+0.03}_{-0.01}$\\
$R^L_{h_c}$  & & $0.06^{+0.02}_{-0.02}$ &$0.08^{+0.02}_{-0.02}$& $0.07^{+0.02}_{-0.02}$\\
  $R^\perp_{h_c}$  & & $0.11^{+0.00}_{-0.01}$ &$0.18^{+0.01}_{-0.01}$& $0.17^{+0.01}_{-0.01}$ \\
  $R_{\chi_{c2}}$  & & $0.04^{+0.00}_{-0.01}$ &$0.05^{+0.01}_{-0.01}$& $0.05^{+0.01}_{-0.01}$ \\
$R^L_{\chi_{c2}}$  & & $0.03\pm0.01$ &$0.04\pm0.01$& $0.04\pm0.01$ \\
  $R^\perp_{\chi_{c2}}$  & & $0.05^{+0.01}_{-0.00}$ &$0.06^{+0.01}_{-0.01}$&$0.07^{+0.01}_{-0.01}$   \\
   \hline \hline
\end{tabular}
\end{center}
\end{table}

The NP effects in $b\to c\tau\nu$ have been explored in a model-independent way in Refs.~\cite{Huang:2018nnq,Alok:2017qsi}, in which the following effective Hamiltonian is introduced:
\begin{eqnarray}
{\cal H}_{\rm eff}&=& \frac{4G_F}{\sqrt{2}} V_{cb}[(1+C_{V_1})O_{V_1} +C_{V_2}O_{V_2}+ C_{S_1}O_{S_1}+C_{S_2}O_{S_2}+C_{T}O_T],
\end{eqnarray}
where
\begin{eqnarray}
O_{V_1} = (\bar c_L\gamma^\mu b_{L})(\bar \tau_{L}\gamma_\mu \nu_{L}), \;\;\;
O_{V_2} = (\bar c_R\gamma^\mu b_{R})(\bar \tau_{L}\gamma_\mu \nu_{L}).
\end{eqnarray}
The scalar and tensor operators  have similar forms except the different Lorentz structures.  The authors have fitted the data and obtained the constraints on the anomalous Wilson coefficients   in different scenarios.
The total $\chi^2/dof$ are at the same level, and thus in the following we will use the consider two scenarios with NP effects in $O_{V_1}$ (denoted as S1) and $O_{V_2}$ (denoted as S2). The fitted  $C_{V_1}$ and $C_{V_2}$ read as:~\cite{Huang:2018nnq}
\begin{eqnarray}
[1+{\rm Re}(V_1)]^2+ {\rm Im}(C_{V_1})^2 =1.27\pm0.06,\;\;\;
C_{V_2}= (0.057+0.573i) \pm (0.050+0.072i).
\end{eqnarray}
For the $O_{V_2}$ contribution to differential decay widths, the form factors defined by axial-vector current must reverse the sign. With these results, we obtain give the NP contributions to the ratios in Tab.~\ref{tab:results2}.

\section{conclusion}
In this paper, we have  investigated the form factors of $B_c$ decays into the S-wave and P-wave charmonium in a model independent way. Unitary constraints combining   Lattice QCD simulations and heavy quark limit constraints  allow  solid theoretical predictions.
The new parametrization form we derived can greatly reduce the hadronic uncertainties. The theoretical prediction for $R_{J/\psi}$  is far below  the LHCb experimental data, which
is challenging the understanding of the fundamental theory of SM.
It is worthwhile  to independently study the helicity-dependent observations such as $R^L_{J/\psi}$  and $R^\perp_{J/\psi}$ in experiments. Besides, other observations $R_H$ for $B_c$ decays into the S-wave and P-wave charmonium are also helpful to check the lepton flavor universality.

Our analyses can be straightforwardly  extended to other $b\to c$ transitions like the $\Lambda_b\to \Lambda_c/\Lambda_c^*$, or  even the doubly bottom baryons $\Xi_{bb}/\Omega_{bb}$ decays into $\Xi_{bc}/\Omega_{bc}$. The latter category  has  small production rates at LHC and smaller branching fractions~\cite{Li:2017ndo,Wang:2017mqp,Wang:2017azm}, making it less impressive at this stage.

\section*{Acknowledgments}
 We thank the discussions with Prof.~Cai-Dian L\"u, Prof.~Ying Li, and Dr.~Zhuo-Ran Huang. This work is
supported in part by   National Natural Science Foundation of
China under Grant No.~11575110, 11655002,  11705092,11735010,  by Natural Science Foundation of Jiangsu under
Grant No.~BK20171471, by the Young Thousand Talents Plan,  by Natural Science Foundation of Shanghai under Grant  No.~15DZ2272100,  by Key
Laboratory for Particle Physics, Astrophysics and Cosmology,
Ministry of Education.

\end{document}